\documentclass[twocolumn,twoside]{CAP-journal}

\title{Dynamic control of intermittent renewable energy fluctuations in two-layer power grids }

\author{ \twlsfb Simona Olmi
    \affiliation{
      Istituto dei Sistemi Complessi\\
      CNR - Consiglio Nazionale delle Ricerche\\
      50019 Sesto Fiorentino, Italy\\
      simona.olmi@fi.isc.cnr.it
    }
    \and \twlsfb Carl H. Totz
    \affiliation{
      Institut f{\"u}r Theoretische Physik\\
      Technische Universit\"at Berlin\\
      10623 Berlin, Germany\\
      carl.hendrik@web.de
    }
   \and \twlsfb Eckehard Sch{\"o}ll
    \affiliation{
      Technische Universit\"at Berlin\\
			and Potsdam Institute\\ 
			for Climate Impact Research\\
      14412 Potsdam, Germany\\
      schoell@physik.tu-berlin.de
    } 
}

\begin{document}

\maketitle

\begin{abstract}
In this work we model the dynamics of power grids in terms of a two-layer network, and use the Italian high voltage power grid as a proof-of-principle example.
The first layer in our model represents the power grid consisting of generators and consumers, while the second layer represents a dynamic communication network that serves as a controller of the first layer. The dynamics of the power grid is modelled by the Kuramoto model with inertia, while the communication layer provides a control signal $P_i^c$ for each generator to improve frequency synchronization within the power grid. We propose different realizations of the communication layer topology and of the control signal, and test the control performances in presence of generators with stochastic power output.
When using a control topology that allows all generators to exchange information, we find that a control scheme aimed to minimize the frequency difference between adjacent nodes operates very efficiently even against the worst scenarios with the strongest perturbations. On the other hand, for a control topology where the generators possess the same communication links
as in the power grid layer, a control scheme aimed at restoring the synchronization frequency in the neighborhood of the controlled node turns out to be more efficient.
\end{abstract}


\section{Introduction}
To reach the goal of limiting the global warming to substantially less than two degrees, integrating renewable and sustainable energy sources into the electrical power grid is essential \cite{UNF15}. Wind and solar power are the most promising contributors to reach a sustainable energy supply but their integration into the existing power transmission and distribution systems remains an enormous challenge \cite{VAC11,JAC11,UEC15}. 
Recently, renewable energy generators, which produce a few kilowatts in the case of residential photovoltaic systems, up to some megawatts in the case of large photovoltaic and wind generators, have become widely dispersed around the world, thus transforming the present power system into a large-scale distributed generation system. 
The drawback of renewable energy power plants is that their output is subject to environmental fluctuations outside of human control, i.e., clouds blocking the sun or lack of wind, and these fluctuations emerge on all timescales displaying non-Gaussian behaviour \cite{HEI10,MIL13,Anvari2016,ANV17a,AUE17}. 
In particular the power grid infrastructure is very critical \cite{WIT16}. 
Due to the design of the current power grid as a centralized system where the electric power flows unidirectionally through transmission and distribution lines from power plants to the customer, the control is concentrated in central locations and only partially in substations, while remote ends, like loads, are almost or totally passive. Therefore it is necessary to design more effective and widely distributed intelligent control embedded in local electricity production, two-way electricity and information flows, thus achieving flexible, efficient, economic, and secure power delivery \cite{Liserre2010,ANV20}.

A Smart Grid \cite{Santacana2010} requires both a complex two-way communication infra\-structure, sustaining power flow between intelligent components, and sophisticated computing and information technologies. 
In particular, control is needed in power networks in order to assure stability and to avoid power breakdowns or cascading failures: one of the most important control goals is the preservation of synchronization within the whole power grid~\cite{COR13,MOT13a,TEG17,CAR20,MOL21,TYL21}. Control mechanisms able to preserve synchronization are ordered by their time scale on which they act: 
the first second of the disturbance is mainly uncontrolled, and in this case a power plant will unexpectedly shut down with a subsequent shortage of power in the system, energy is drawn from the spinning reserve of the generators. Within the next seconds, the primary control sets on to stabilize the frequency and to prevent a large drop. Finally, to restore the frequency back to its nominal value of 50 (or 60) Hertz, secondary control is necessary.  In many recent studies on power system dynamics and stability, the effects of control are completely neglected or only primary control is considered \cite{Dorfler13,SCH18c,ROH12,SCH15,SCH16,WANG16}. This control becomes less feasible if the percentage of renewable power plants increases, due to their reduced inertia \cite{ULB14,DOH10}. Few studies are devoted to secondary control \cite{Weitenberg18,Tchuisseu18,Simpson12} and to time-delayed feedback control \cite{Okuno2006,Dongmo17,TAH19}. 

The aim of this work is to develop novel control concepts considering the communication infrastructure present in the smart grid. 
In applied nonlinear dynamics few works have included the communication layer into the modeling of power networks. Even though the communication infrastructure plays an important role in control and synchronization, preliminary works \cite{Li11,Wei12} assume trivial networks, without disconnected nodes, isolated generators, microgrids, or even coupled microgrids that can be connected or disconnected to the main grid at any time. 

In this paper we consider a two-layer network in a full dynamic description \cite{TOT20}. It consists of a power grid layer and a communication layer, which provides the control for the power grid.
Each layer is governed by its own dynamics, which is dependent upon the state of the other layer. In particular the physical topology that relates the interconnection of distributed generators and loads is described by coupled Kuramoto phase oscillators with inertia, derived from the swing equation \cite{Filatrella:EPJB:61}, while 
the communication topology, which describes the information flow of the power system control measurements, depends on the information of the neighbors of each node \cite{Giraldo:CDC:2013}.
Starting from the ideal synchronized state, we investigate the effect of a real threat to synchronization of the network: intermittent noise, which is used to describe the fluctuating power
output of renewable energy power plants. For this perturbation different setups of the communication layer are tested to find an effective control strategy that successfully preserves frequency synchronization. 
As a proof of concept the Italian high voltage power grid is considered. The same two-layer topology has been investigated in \cite{BUL10} to understand how localized events can present a severe danger to the stability of the whole power grid by causing a cascade of failures, but without considering the dynamics of the control nodes. 
In \cite{TOT20} we have investigated, within the same set-up, different perturbations to which the system is subject, e.g., failure of nodes, increased consumer demand, power plants subject to non-Gaussian or Gaussian white noise. Here we restrict ourselves to a more detailed analysis of intermittent power fluctuations, typical for renewable energy sources.
Our proposed control techniques preserve synchronization for the implemented perturbations, thus demonstrating the powerful perspectives of our control approach which considers synchronization of power systems based on the coupled dynamics of the smart grid architecture and the communication infrastructure.

\section{Model and Methods}

\begin{figure*}[ht]
		\includegraphics[width=0.32\linewidth]{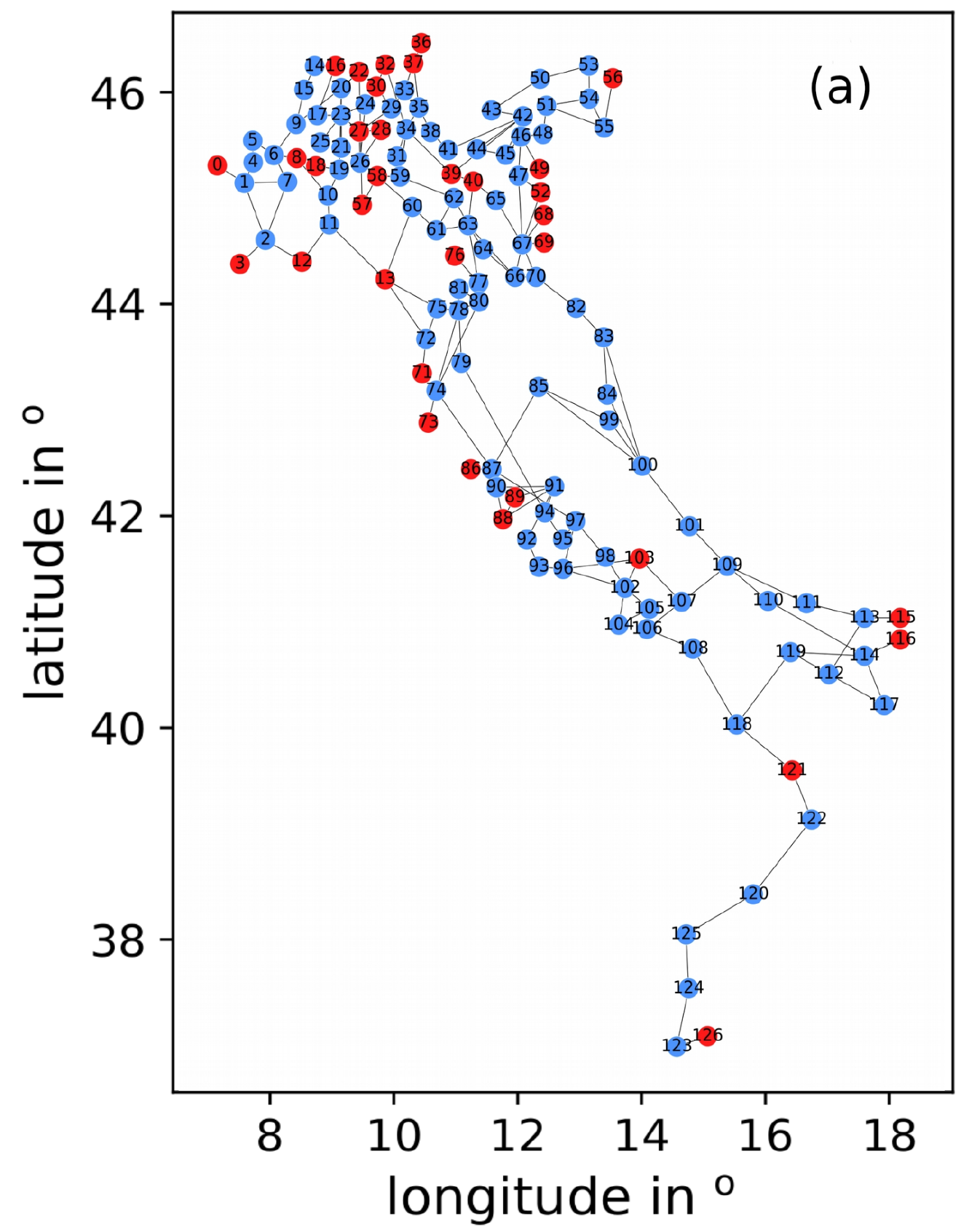}
		\includegraphics[width=0.32\linewidth]{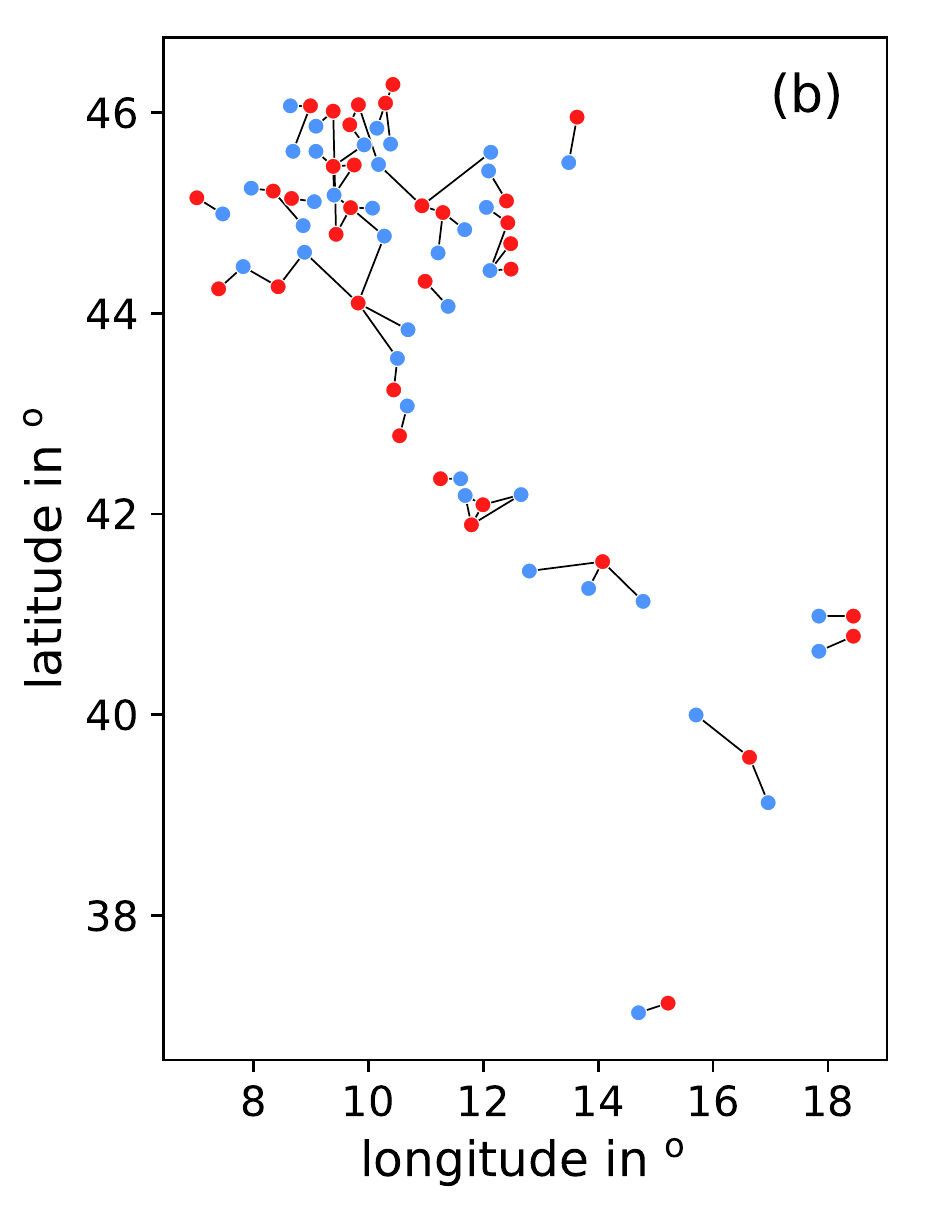}
		\includegraphics[width=0.32\linewidth]{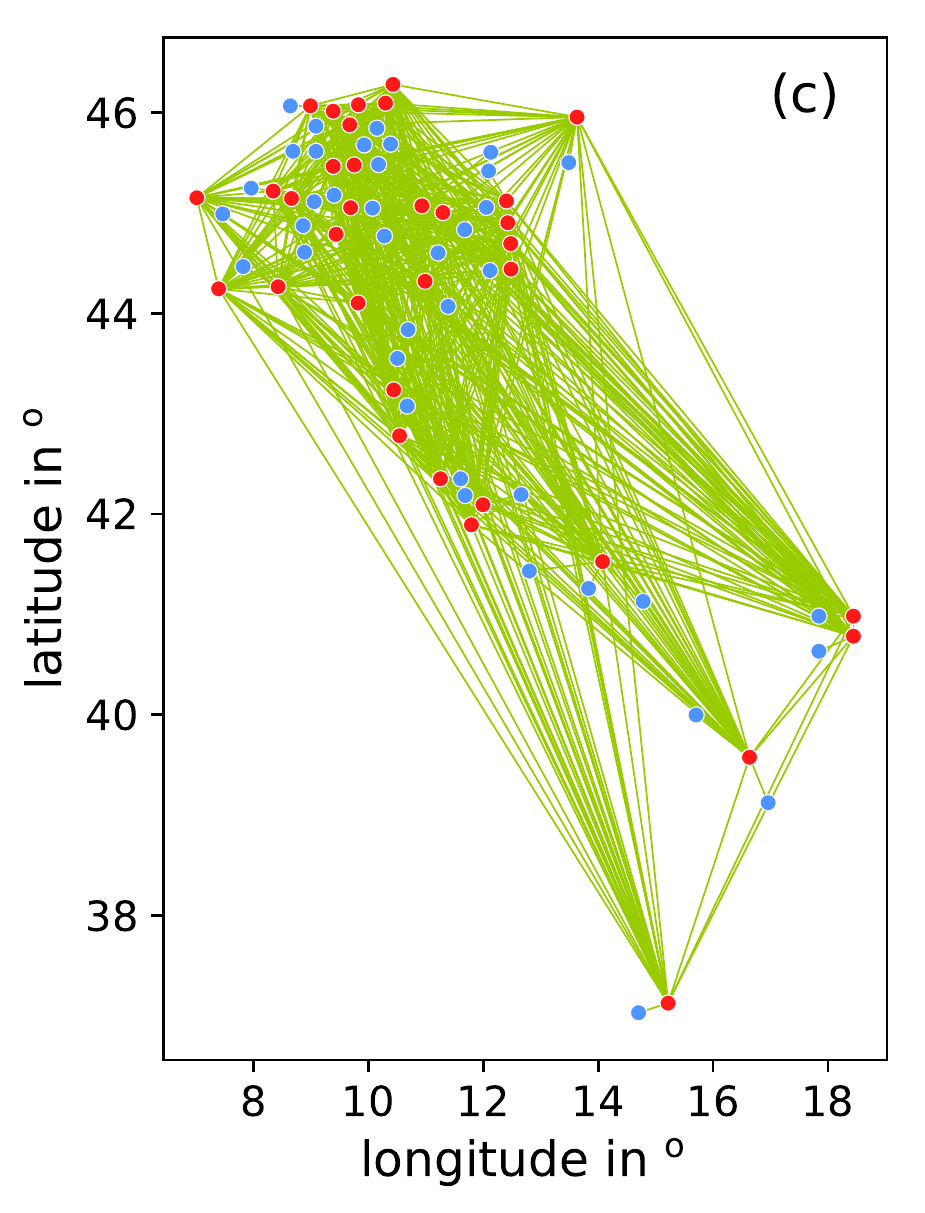}
	\caption{Visualisation of the topology of the individual layers of the two-layer power grid:
		(a) Topology $a_{ij}$ of the power grid layer based on the real Italian high voltage power grid;
		(b) control layer topology $c_{ij}^{loc}$ where the communication links of the generators are as in the power grid layer;
		(c) control layer topology $c_{ij}^{ext}$ where generators possess additional communication links to all other generators in the network (green). Red nodes denote generators, while blue nodes denote consumers. 
Position of nodes has been slightly modified to improve readability.
	}
	\label{fig:topologies}
\end{figure*}

\textbf{Power grid layer.} The Kuramoto model with inertia describes the phase and frequency dynamics of $N$ coupled synchronous machines arranged in the controlled power grid layer, i.e., generators or consumers within the power grid, where mechanical and electrical phase and frequency are assumed to be identical:  
\begin{eqnarray}\nonumber
&& m\ddot{\vartheta_i}(t)= -\dot{\vartheta_i}(t) +\left(\Omega_i+P_i^c(t)\right) \\\label{dynamics-eq}
&+ &K\sum_{j}^N a_{ij}(t)\, \sin\left(\vartheta_j(t)-\vartheta_i(t)\right) \,,
\end{eqnarray}
with the phase $\vartheta_i$ and frequencies $\dot{\vartheta}_i$ of node $i=1,...,N$. Both dynamical variables $\vartheta_i$, $\dot{\vartheta}_i$ are defined relative to a frame 
rotating with the reference power line frequency (i.e., $50$ or $60 \, \text{Hz}$). The inherent frequency distribution is bimodal, where a positive natural frequency $\Omega_i$ of a node corresponds to the suitably normalized power supplied by a generator, while a negative natural frequency corresponds to the demand of a load.
The power balance requires that the power supplied by all generators in the network is exactly met by the combined demand of all loads: $\sum_i \Omega_i=0$.
The additional term $P_i^c$ denotes the control signal supplied by the communication layer, which serves as an offset to the power supplied by a controlled generator.
For simplicity we assume homogeneously distributed transmission capacities $K$ and inertia $m$.
The adjacency matrix $a_{ij}$ takes the value 1 if node $i$ has a transmission line connected to node $j$, 0 otherwise. In our numerical simulations we use the Italian high voltage power grid topology \cite{GEN19}, which consists of $N=127$ nodes, of which $34$ are generators and $93$ are loads. The matrix $a_{ij}$, which describes the topology, is unweighted and symmetrical (see Fig.\thinspace\ref{fig:topologies}\thinspace (a) for graph details). 

\textbf{Macroscopic indicators and parameter regime.} For our investigation we have chosen a regime of bistability in which both the fully frequency-synchronized state and a partially synchronized state are accessible. In this way it is possible to mimic the effect of a perturbation applied to the synchronized state: this displaces the system out of synchrony into an intermediate state where the operating conditions are not optimal for the functioning of power grids.
For a detailed analysis of the system in absence of control see \cite{olmi2014,TUM18,TOT20}. For each numerical experiment, the system evolves for a transient time $t_R$, until the network settles and it reaches a steady state. At this point, characteristic measures are calculated, averaging over a time $t_A$, in order to assess the level of synchronization of the state $\{\vartheta_i(t_R), \dot{\vartheta}_i(t_R)\}$. In particular the time-averaged phase velocity profile $\langle\omega_i\rangle_t\equiv\langle\dot{\vartheta}_i\rangle_t$ provides information about the frequency synchronization of each node, while the standard deviation of the frequencies 
\begin{equation}\label{eq:average_stdev}
 \Delta\omega(t)=\sqrt{\frac{1}{N}\sum_{j}^{N}(\omega_j(t)-\overline{\omega}(t))^2}
\end{equation}
gives information about the deviation from complete frequency synchronization at the macroscopic level ($\bar{\omega}(t)$ represents the instantaneous ensemble-averaged grid frequency). 
The phase ordering of the power grid is measured by the complex order parameter
\begin{equation}\label{EQ:order_parameter}
R(t) e^{i\Phi(t)} = \frac{1}{N} \sum\limits_{j=1}^N e^{i\vartheta_j},
\end{equation}
where the modulus $ R(t) \in \left [ 0 , 1 \right ] $ and the argument $ \Phi(t) $ indicate the degree of synchrony and mean phase angle, respectively. 
In the following we will denote $ R(t) $ as \textit{global order parameter}. In the continuum limit an asynchronous state is characterized by $R \approx 0$, while $R=1$ 
corresponds to full phase synchronization. Intermediate values of $R$ correspond to states with partial or cluster synchronization. 
In this article we explore the dynamics of the system at $K=6.5$, where the system shows bistability between full frequency synchronization (i.e., it corresponds to the minimum coupling strength for which full frequency synchronization is still achievable) and partial synchronization ($\langle \Delta\omega\rangle_t\simeq 1.6$), which models the resulting state when the power grid is strongly perturbed.

\textbf{Communication layer.} The smart grid includes a communication infrastructure in all the stages of the power system, from transmission to users, allowing for the design of control strategies based on the information data flow. 
In real applications, we need to consider isolated elements, where synchronization needs to be assured such that, if the isolated nodes are reconnected to the main power system, failures can be avoided and the stability of the network is preserved. Therefore, the use of a communication layer of the network may improve the performance of the power system. We consider two layer topologies or infrastructures, the physical topology that describes the power system dynamics (as shown in Fig.\thinspace\ref{fig:topologies}\thinspace (a)), and  the communication layer topology, which describes how data from each node is transmitted (see Fig.\thinspace\ref{fig:topologies}\thinspace (b), (c) for the topologies of the communication layer investigated here). Both infrastructures can be conceived as a multilayer network.

To design a control strategy for synchronization, it is necessary to collect information from each generator and its neighbors. Phasor measurement units or sensors provide information, such that local controllers integrated with the generator nodes use the information to calculate a control signal $P^c_i\in \mathbb{R}$. The loads are not controlled. The control signal can be interpreted as power injection for positive $P^c_i$ or power absorption for negative values of $P^c_i$, which is realized using storage devices (e.g., batteries) that can  absorb or inject power to the generator buses \cite{qian2010}. This real-world framework can be translated in terms of Eq. (\ref{dynamics-eq}) as injecting power in steady state operation.
 
Since the communication layer describes the exchange of information between the nodes about their current dynamic states, we consider a control strategy that depends on the information of the neighbors of each node, from which a control signal $P_i^c(t)$ for each controlled node is calculated dynamically. 
Neighbors can be related using the adjacency matrix $\mathcal{C}=\{c_{ij}\}$ of the communication layer. Our essential point is that we equip the communication layer also with a dynamics of its own. Therefore we propose to determine the control signal $P_i^c(t)$ by a first order differential equation depending on the frequencies $\dot{\vartheta_j}$ of neighboring nodes within the communication layer $c_{ij}$:
\begin{align} \label{control_general}
\dot{P_i^c}(t)=
G_i\,f_i\left(c_{ij},\{\dot{\vartheta_j}(t)\}\right) \,,
\end{align}
where $G_i$ is the control strength and $f_i$ represents the control function. In particular we assume that it is possible to control only the power output of generators in the network, thus $G_i$ is zero for all loads: 
\begin{align} \label{G_i_eq}
G_i= 
\begin{cases}
G \,\text{,} &i\in M_{gen}\\
0 \,\text{,} &\text{otherwise}
\end{cases}
\,
\end{align}
where $M_{gen}$ is the set of all generators in the network. Throughout this work we choose $G=0.04$.
Two different topologies have been considered for the communication layer, namely $c_{ij}^{loc}$ and $c_{ij}^{ext}$.
In the local topology $c_{ij}^{loc}$ the connections between each generator and the other nodes in the communication layer correspond to the connections in the power grid layer (i.e., the communication layer network consists of a subnetwork of the power grid layer which contains all links of the generators in the power grid), except that each node has available also the information about itself, so that the diagonal elements of the adjacency matrix are nonzero.
The local topology is thus described by the adjacency matrix
\begin{align} \label{cij_loc_def}
c_{ij}^{loc}= 
\begin{cases}
1 \,\text{,} &i=j\\
a_{ij} \,\text{,} &\text{otherwise}
\end{cases}
\,.
\end{align}
As only generators receive a control signal $P_i^c$, all loads which are not connected to a generator can be disregarded in the communication layer, as illustrated in Fig.\thinspace\ref{fig:topologies}\thinspace (b).

For the extended control topology $c_{ij}^{ext}$ additional links between all generators are present (see Fig.\thinspace\ref{fig:topologies}\thinspace (c)). The globally coupled generators represent a subset of the communication layer (which additionally contains the neighboring loads); this provides an exchange of information between all generator control stations. The corresponding connectivity matrix is defined as
\begin{align} \label{cij_ext_def}
c_{ij}^{ext}= 
\begin{cases}
1 \,\text{,} &i\wedge j\in M_{gen}\\
c_{ij}^{loc} \,\text{,} &\text{otherwise}
\end{cases}
\,.
\end{align}
As shown in Eq. \eqref{control_general}, the dynamics of the control signal is governed by the control function $f_i\left(c_{ij},\{\dot{\vartheta_j}\}\right)$.
In this work we consider three different control functions: the first one is to apply a control signal that is proportional to the frequency difference between node $i$ and its neighbors \cite{Giraldo:CDC:2013}
\begin{align} \label{diffC-eq}
f_i^{diff}\left(c_{ij},\{\dot{\vartheta_j}(t)\}\right)=
\sum_j^N c_{ij}\left(\dot{\vartheta_j}(t)-\dot{\vartheta_i}(t)\right)
\,.
\end{align}
We refer to this control scheme as \textit{difference control}. In control theory this control function is known as proportional control \cite{Bequette2003}. However, when
considering a control proportional to the frequency error difference, we can also refer to it as frequency droop control \cite{Brabandere2007} or diffusive coupling \cite{Hale1997}, depending on the addressed community (i.e., power systems control or circuit theory, respectively).
\\
The second approach is to apply a control signal that aims to restore power balance in the neighborhood of the controlled node:
\begin{align} \label{dirC-eq}
f_i^{dir}\left(c_{ij},\{\dot{\vartheta_j}(t)\}\right)=
-\frac{1}{N_i}\sum_j^N c_{ij}\,\dot{\vartheta_j}(t)
\,.
\end{align}
Here $N_i$ gives the number of direct neighbors of node $i$ in $c_{ij}$. We will call this control scheme \textit{direct control}. A control scheme proportional to the absolute frequency error is referred to as proportional control in control theory \cite{Bequette2003}.
\\
The final control scheme is a combination of both difference and direct control:
\begin{eqnarray}            \label{bothC-eq}
\nonumber
f_i^{comb}\left(c_{ij},\{\dot{\vartheta_j}(t)\}\right)&=&
\sum_j^N c_{ij}\left(a\left(\dot{\vartheta_j}(t)-\dot{\vartheta_i}(t)\right)\right. \\
&-& \left. b\frac{\dot{\vartheta_j}(t)}{N_i}
\right)
\,.
\end{eqnarray}
Here $a$ and $b$ are weight factors of the two components. 
In all further instances we will assume that $a=b=1$. 
We will refer to this control scheme as \textit{combined control}.

\textbf{Perturbation: Generation of intermittent noise.}
Due to atmospheric turbulence, wind power has specific turbulent characteristics \cite{MIL13,Anvari2016,ANV17a,AUE17}, such as extreme events, time correlations, Kolmogorov power spectrum, and intermittent increment statistics. In particular the increment probability density functions of real wind power data significantly deviate from Gaussianity and its power spectrum displays $(5/3)$-decay with some discrepancy in the high frequency range. Based on this, we generate intermittent power time series $x(t)$ according to the synthetique feed-in noise generation detailed by Schmietendorf et al. in \cite{Schmietendorf:EPJB:90}. The first step in generating the intermittent noise time series is to consider the dynamics of the following Langevin-type system of equations:
\begin{align} \label{interm_noise_base}
\dot{z}(t)&=z(t)\left( g-\frac{z(t)}{z_0}\right) +\sqrt{Iz^2(t)}y(t)
\,,
\\
\dot{y}(t)&=-\gamma y(t) +\xi(t)
\,,
\end{align}
where $y(t)$ represents colored noise generated by an Ornstein–Uhlenbeck process \cite{GIL96,RIC88} with a correlation time $\tau_{OU}=1/\gamma$ and with a $\delta$-correlated Gaussian white noise term $\xi$.
The parameter $I$ controls the intermittency strength, while the other parameters $\gamma = 1.0$, $g = 0.5$ and $z_0 = 2.0$ are chosen as in \cite{Schmietendorf:EPJB:90}.
In a second step the time series $z(t)$ is transformed, so that its power spectrum resembles more closely the power spectrum of wind power plants.
To achieve this, the Fourier transform
$X(f)=FT\left[ z(t) \right] (f)$
is first divided by its amplitude spectrum.
This process eliminates the amplitude information of $X(f)$, but retains its phase information.
Subsequently a weight function $h(f)$ is used in order to make the spectrum of the series similar to the empirical data:
\begin{align} \label{IN_generation_2}
\hat{X}(f)=\frac{X(f)}{\left| X(f)\right|} h^{\frac{1}{2}}(f)
\,.
\end{align}
The power spectrum of $\hat{X}(f)$ is proportional to the weight function.
Finally $\hat{X}$ is transformed back into the time domain: $\tilde{x}(t)=FT^{-1}[\hat{X}(f)](t)$.
Due to the elimination of amplitude information, the amplitude of $\tilde{x}$ is freely scalable. 
The standard deviation $\sigma_{\tilde{x}}$ of the aggregated distribution of $\tilde{x}$ is rescaled to any desired $\sigma_{{x}}$:
\begin{align} \label{IN_final}
{x}(t)=
\frac{\sigma_{{x}}}{\sigma_{\tilde{x}}}\tilde{x}(t)
\,.
\end{align}
Since $\langle x(t)\rangle = 0$, $\langle\Omega_i(t)\rangle=\Omega_{gen}$ and power balance is maintained on long-time average.

Further restrictions are introduced to make this perturbation more realistic.
A lower boundary for ${x}$ is introduced, so that a generator cannot operate as a load in the network due to the influence of noise: all values $x<-1$ are truncated to $x=-1$.
Furthermore a power feed-in cut-off is assumed, which means that generators have a maximum power output they can supply:
all values $x>1$ are truncated to $x=1$.
This additionally truncates some of the extreme events in the strongly intermittent power time series, while the mean and standard deviation are nonetheless almost unaffected by this.
With these constraints for ${x}$, the penetration parameter $\mu$ is chosen equal to the natural frequency $\Omega_i$ of the affected generator to prevent any generator from acting as a load and giving a maximum power feed-in to the network.
\\
Throughout this work we fix $I=2$,  $h(f)=f^{-\frac{5}{3}}$ and $\sigma_x=\frac{1}{3}$ for intermittent noise.

\section{Results}

A characteristic feature of renewable energy sources are power fluctuations due to fluctuating wind and solar irradiation (clouds).
This requires novel design concepts and theoretical investigations into smart storage control strategies to balance feed-in variations and mitigate power quality problems induced by stochastic fluctuations. A particular challenge for stable power grid operation is given by wind- and solar-induced fluctuations, which follow characteristic non-Gaussian statistics over a broad band of time scales from seasonal and diurnal imbalances down to short-term fluctuations on the scale of seconds and sub-seconds \cite{Anvari2016}. The turbulent character of wind feed-in, and in particular its intermittency, is directly transferred into frequency and voltage fluctuations, as shown in \cite{Schmietendorf:EPJB:90}, where the main characteristics of real wind feed-in were captured by generating intermittent time series on the basis of a Langevin-type model and imposing a realistic power spectrum.
In the following we systematically investigate the effect of applying intermittent noise to each individual generator (\textit{single node perturbation}).

\begin{figure}[t]
	\includegraphics[width=0.51\textwidth]{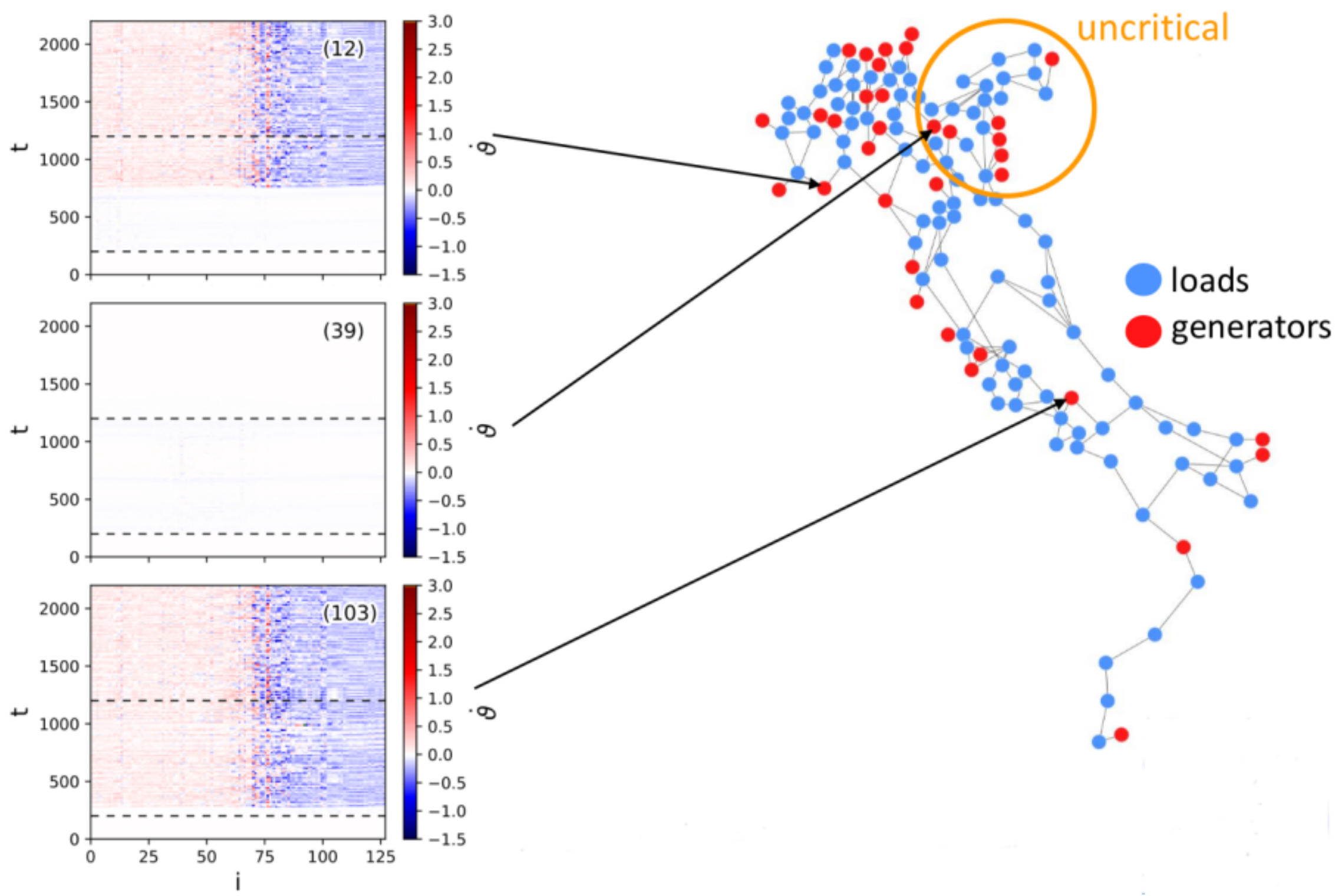}
	\caption{Frequency synchronization of the network is vulnerable to disruption of nodes depending on their location: 
	Applying intermittent noise to any generator outside the orange circle is critical to frequency synchronization. Generators are marked by red dots, while blue dots indicate loads. Space-time plots of the network, for 3 different perturbed nodes, are shown in the left column. Color indicates the instantaneous frequency $\dot{\vartheta}$ of the nodes, where the horizontal dashed lines mark the onset and the end of the perturbation. Each panel is labeled by the number of the perturbed node in brackets. Coupling strength $K=6.5$,
		inertial mass $m=10$, 
		bimodal frequency (power) distribution $\Omega_{load}=-1$ and $\Omega_{gen}=\frac{93}{34}$.
		Integration time step $\Delta t=0.002$. 
	}
	\label{fig:perturbation_pattern}
\end{figure}

\begin{figure*}[t]
	\includegraphics[width=0.49\textwidth]{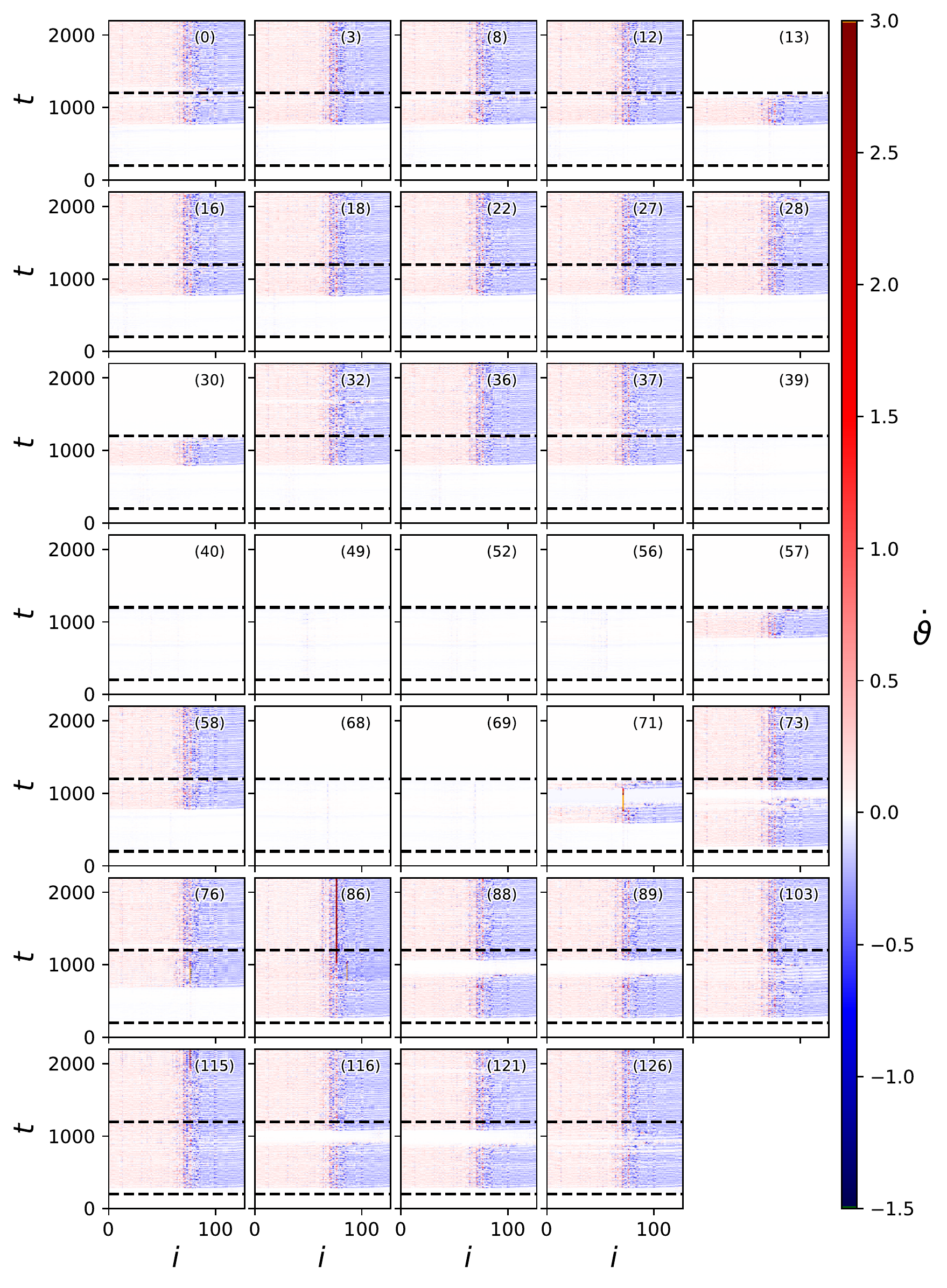}
	\includegraphics[width=0.49\textwidth]{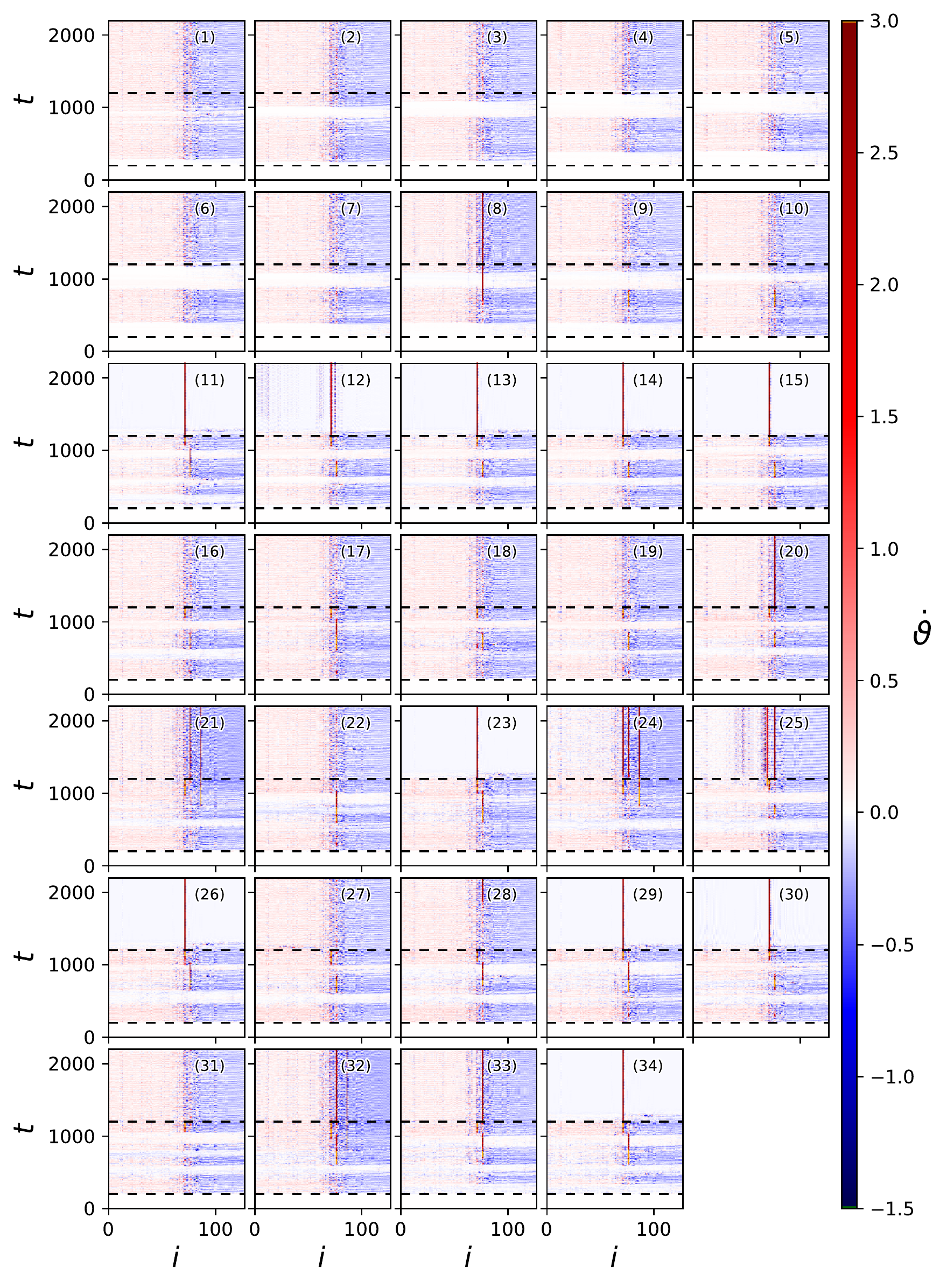}
	\caption{
		Space-time plots of the network, when a single generator (left) or multiple generators (right) are subjected to intermittent noise with $\mu=\frac{93}{34}$.
		Node index $i$ on the x-axis and time $t$ on the y-axis. Left panels: the index $i$ of the affected generator is noted in the upper right corner of each panel. Right panels:  The number $n$ of affected generators is noted in the upper right corner of each panel. Dashed horizontal lines indicate the starting time $t_{start}$ and ending time $t_{end}$ of the perturbation, respectively. Parameters $\mu=\Omega_{gen}$, $\sigma_x=\frac{1}{3}$, $I=2$, $h(f)=f^{-\frac{5}{3}}$, $g=0.5$, $z_0=2$, $\gamma=1$.	
	}
	\label{space_time_plot_nocontrol}
\end{figure*}

\begin{figure}[t]
		\includegraphics[width=1.\linewidth]{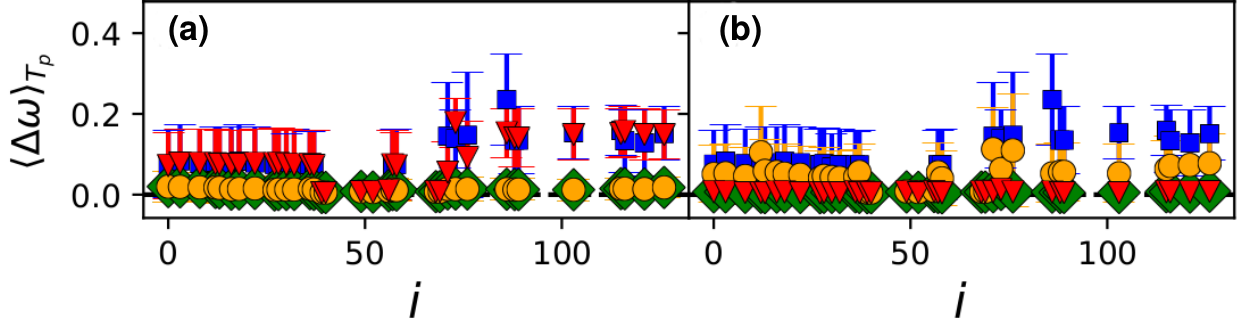}
	\caption{Control of frequency synchronization for intermittent noise targeting a single node: Mean standard deviation of frequency during the time of perturbation $\langle\Delta\omega\rangle_{T_p}$ vs the index $i$ of the targeted node. 
	The symbols indicate different control schemes. Blue squares: no control. Red triangles: \textit{difference control}. Yellow circles: \textit{direct control}. Green diamonds: \textit{combined control}.
	Control strength: $G=0.04$.
	The two panels show different control layer topologies: The left panel (a) shows the simple setup ($c_{ij}^{loc}$) of the control layer, where generators possess the same connections as in the power grid layer, and the right panel (b) shows a control layer topology ($c_{ij}^{ext}$) where additional links between all generators are present.
	Parameters as in Fig. \ref{space_time_plot_nocontrol}.
	}
	\label{fig:CMode_Comp_single}
\end{figure}

When the perturbation is applied systematically to each individual generator in the network in the absence of control, we observe that the whole network always loses synchronization, irrespectively of the targeted node (Figs. \ref{fig:perturbation_pattern},\ref{space_time_plot_nocontrol}). Only few generators in the north-east of Italy are resilient to perturbation, as exemplified in Fig. \ref{fig:perturbation_pattern} (middle panel).  Due to the perturbation, the frequencies of the southern part of the Italian grid (index $i>70$) usually start fluctuating with slightly negative frequencies, while the frequencies of the northern part fluctuate correspondingly with slightly positive frequencies. This behavior emerges during the perturbation and remains even after the perturbation ends. However generators do not usually desynchronize from the rest of the power grid, apart from a few cases (i.e., nodes $i$=86, 115), which correspond to dead ends. Moreover fluctuations affect the network with a different timing, depending on the topological location: if a node belonging to the southern part of the network is targeted, the network loses synchronization almost immediately; if a node belonging to the north-western part of the power grid is targeted, the power grid eventually loses frequency synchronization after some time $t\sim T_p/2$ where $T_p$ is the duration of the perturbation.
Only the generators in the north-eastern part of the power grid ($i=39,\,40,\,49,\,52,\,56,\,68,\,69$) remain resilient as highlighted in Fig.\ref{fig:perturbation_pattern} by an orange circle. This also shows up in the left-hand side of Fig.\thinspace \ref{space_time_plot_nocontrol}, which presents the space-time plots of the instantaneous frequencies if generator $(i)$, whose number is given in the legend of the respective panel, is perturbed (left panels). 

\begin{figure*}[t]
	\includegraphics[width=0.33\linewidth]{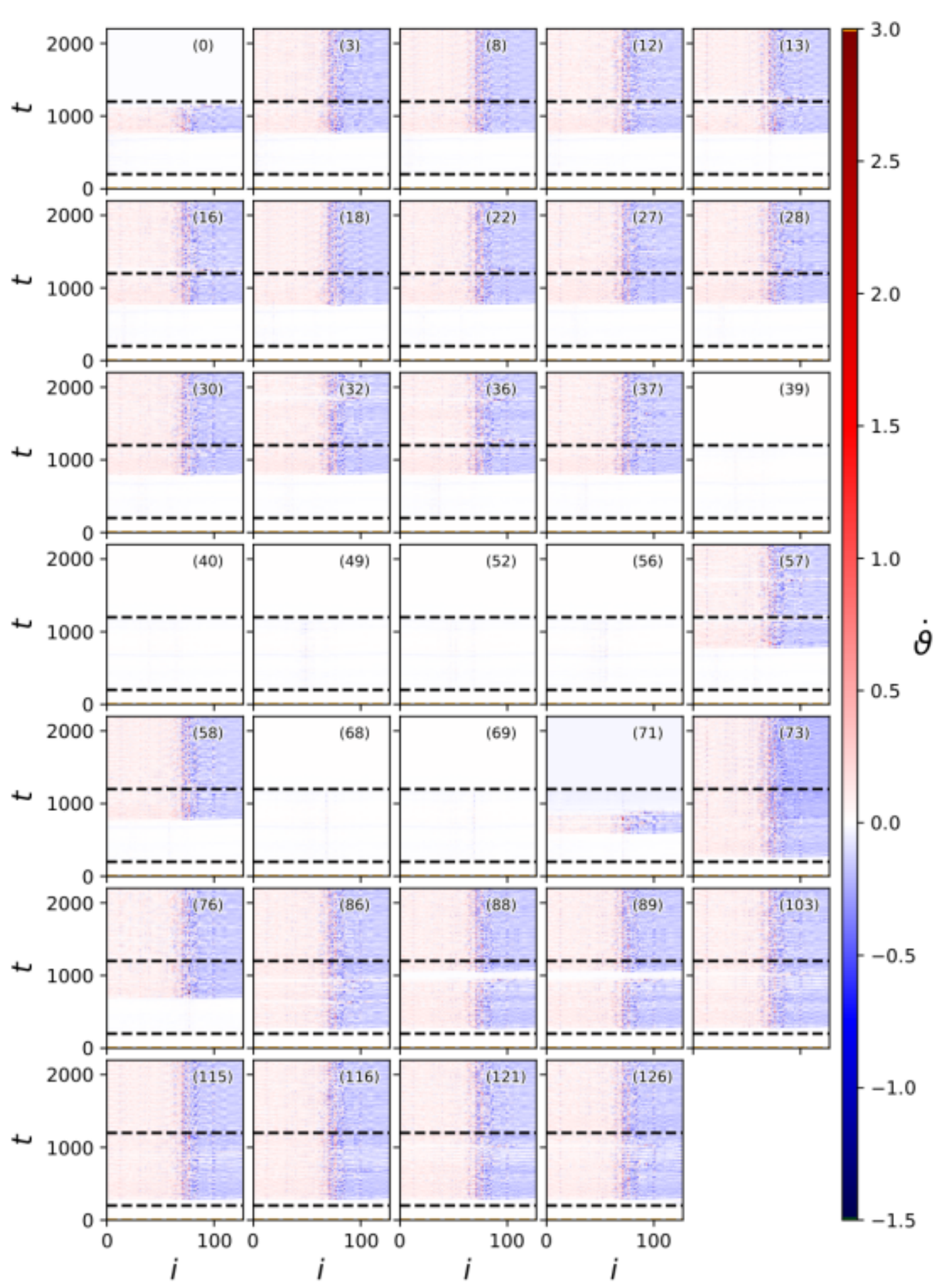}
	\includegraphics[width=0.33\linewidth]{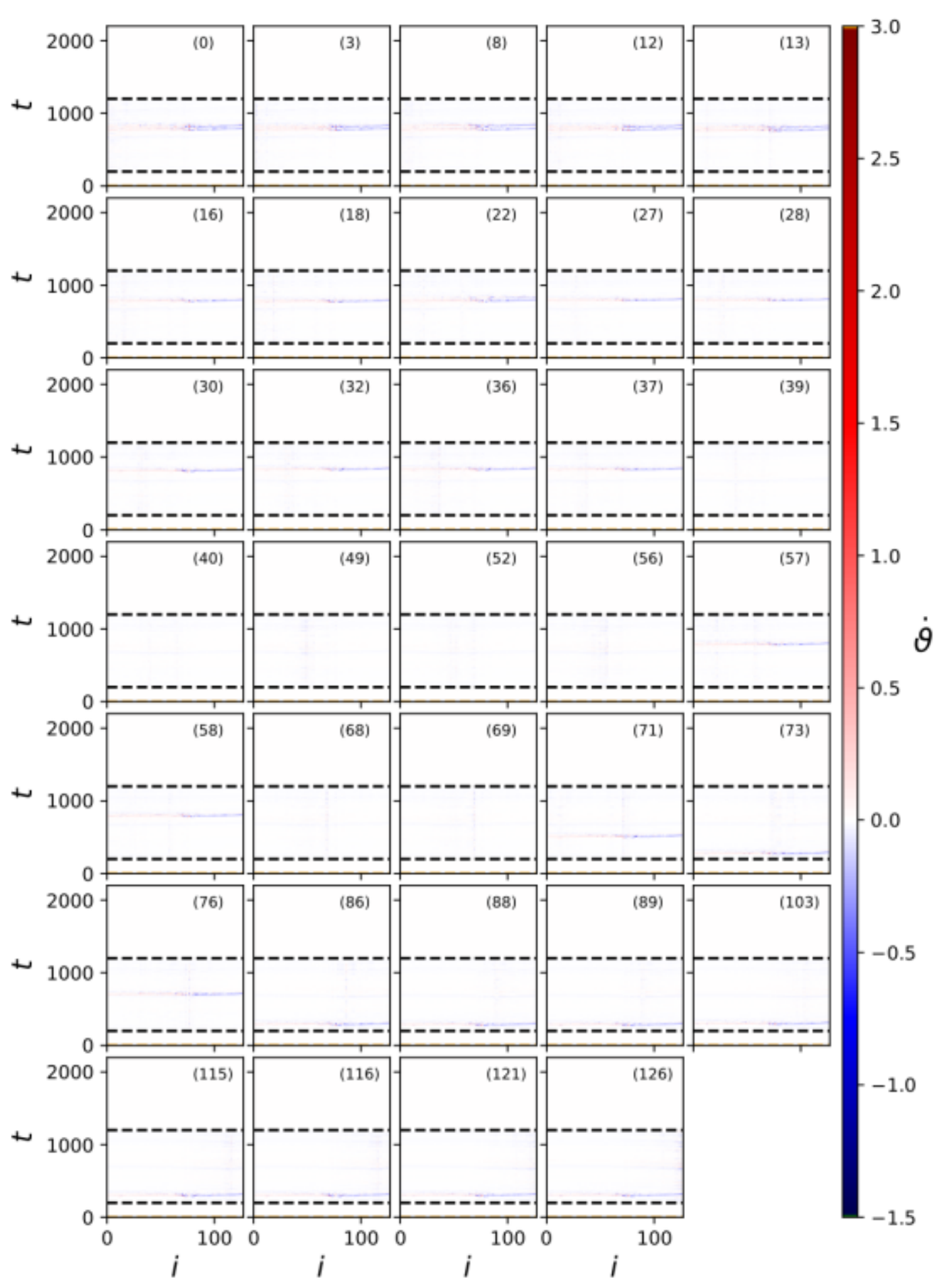}
	\includegraphics[width=0.33\linewidth]{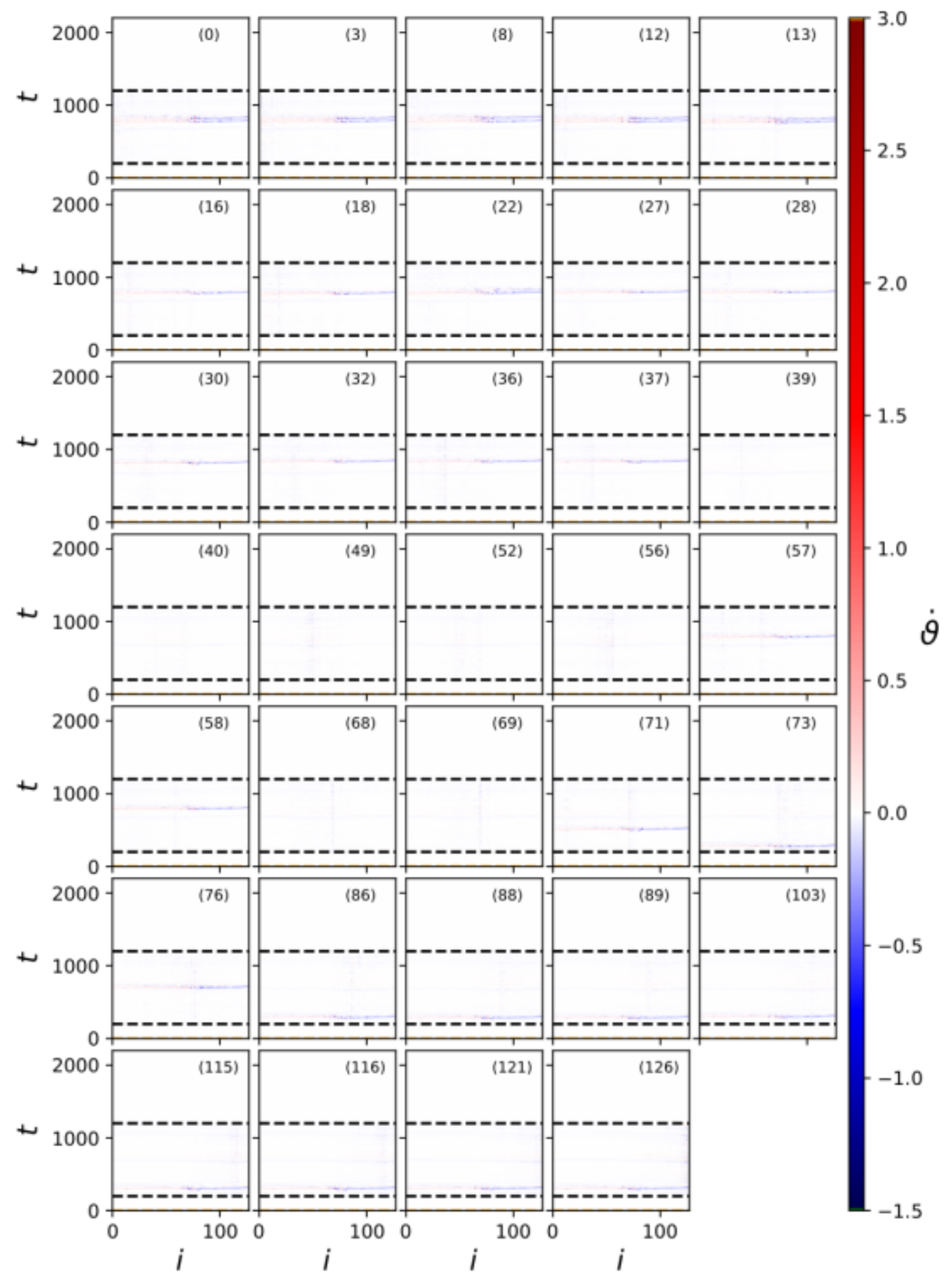}
	\caption{Space-time plots of the network, when a single generator is subject to intermittent noise. Generators are controlled by difference-control (left), direct-control (center), combination-control (right) in the local control topology $c_{ij}^{loc}$. Node index $i$ on the x-axis and time $t$ on the y-axis. Color indicates the frequency $\dot{\vartheta}$. The index $i$ of the affected generator is noted in the upper right corner of each panel. Dashed horizontal lines indicate $t_{start}$ and $t_{end}$ respectively. Parameters as in Fig. \ref{space_time_plot_nocontrol}.
	}
	\label{space_time_plot_singlecontrol_local}
\end{figure*}

\begin{figure*}[t]
	\includegraphics[width=0.33\linewidth]{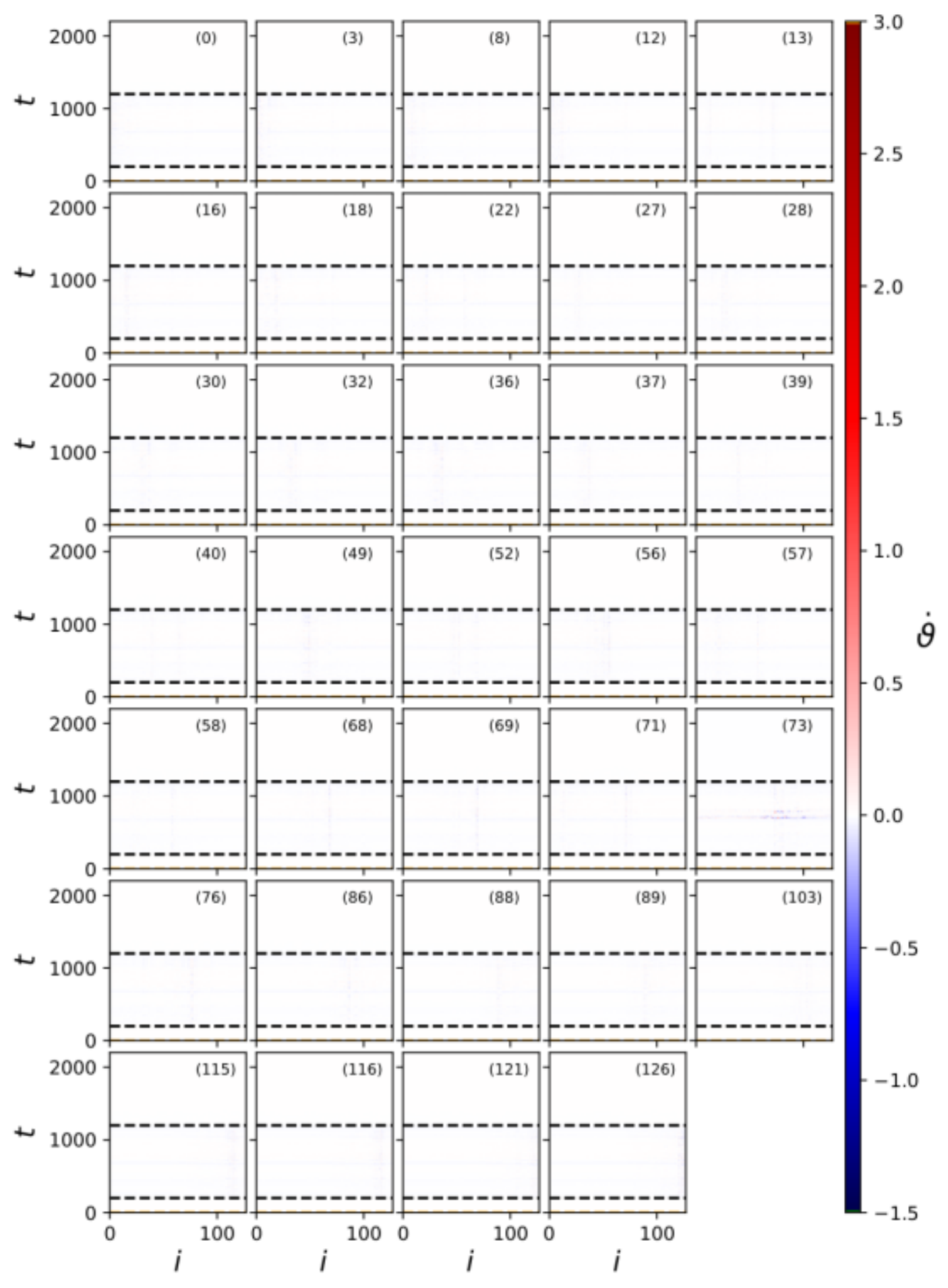}
	\includegraphics[width=0.33\linewidth]{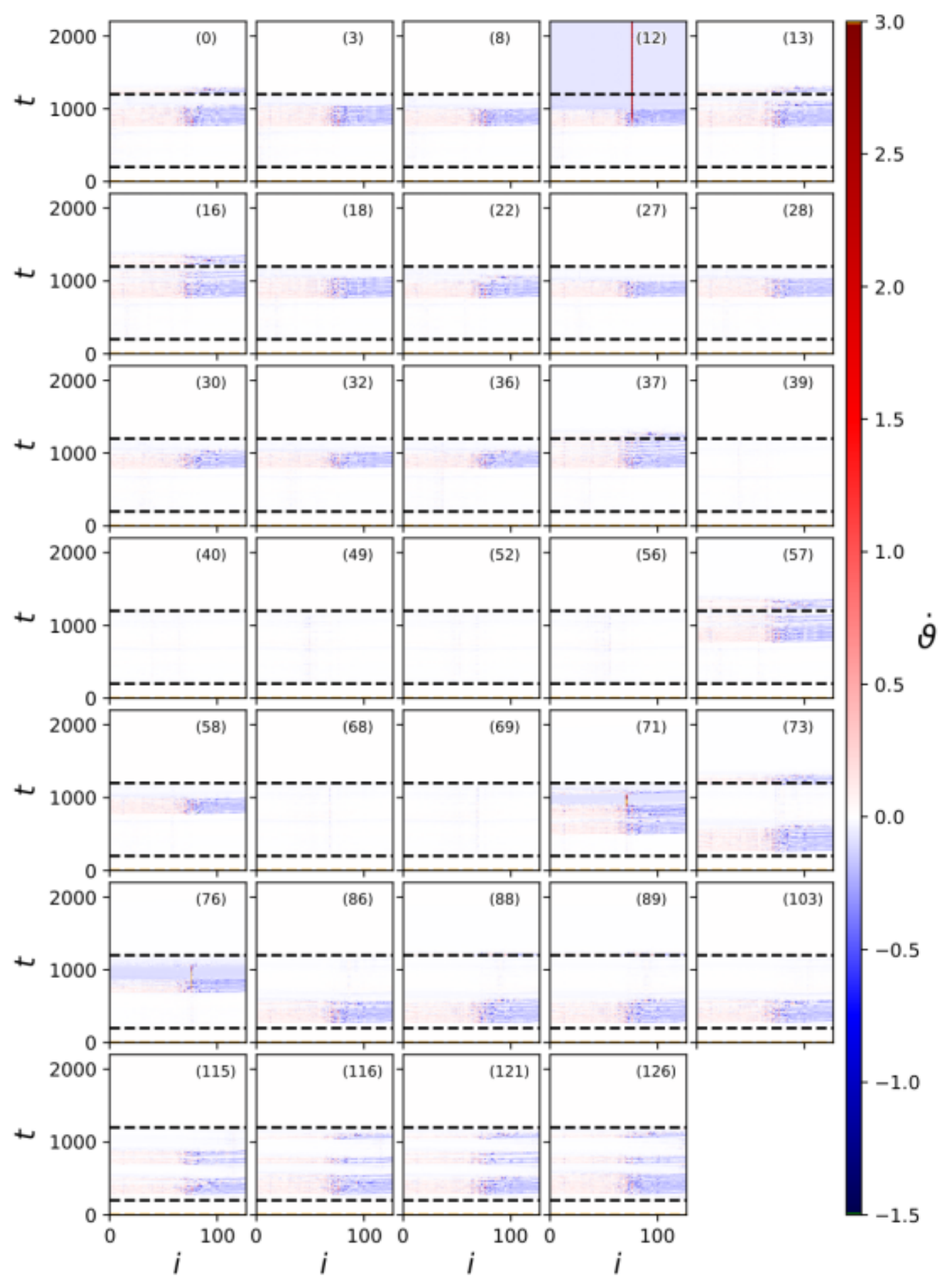}
	\includegraphics[width=0.33\linewidth]{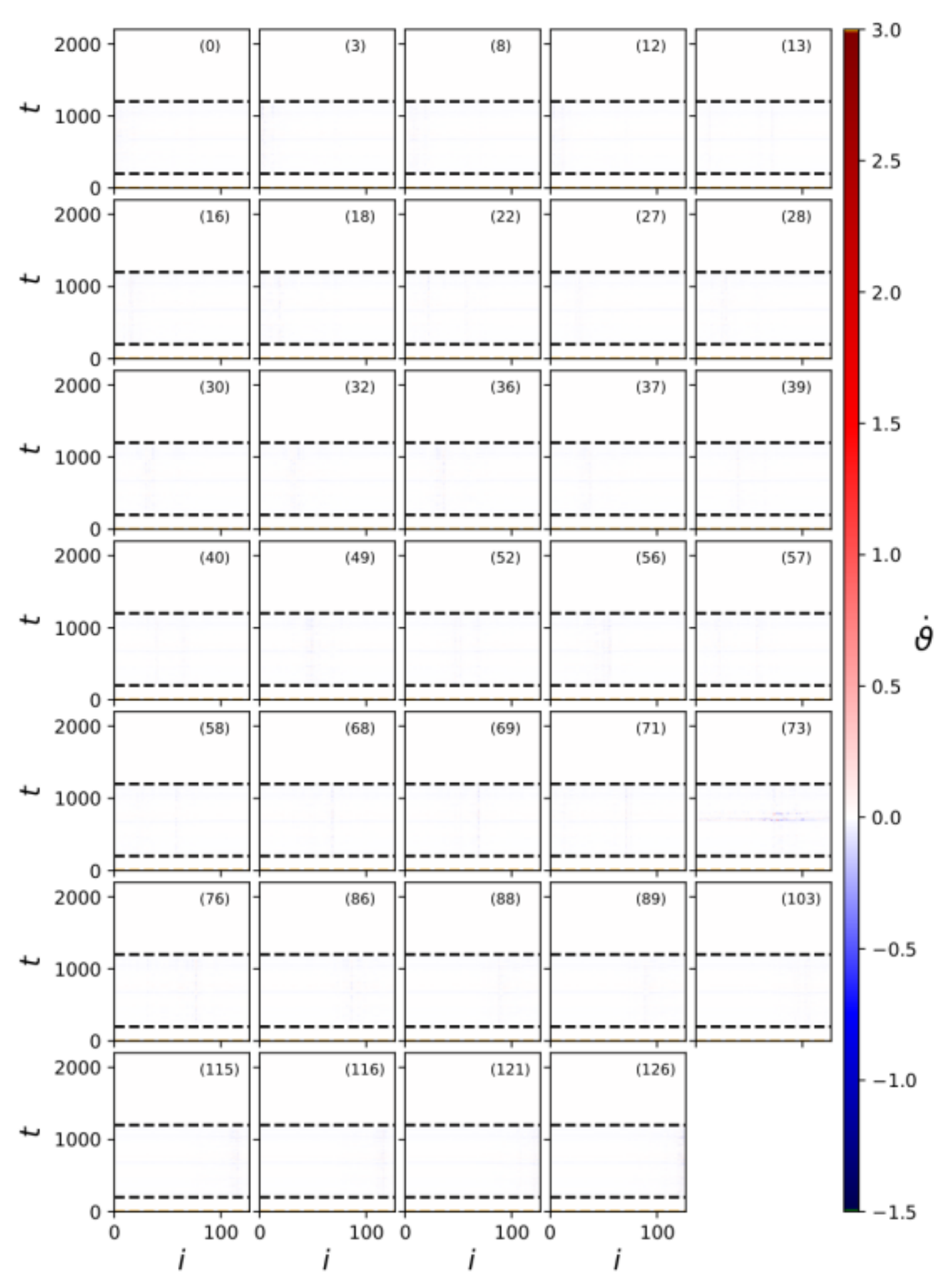}
	\caption{Space-time plots of the network, when a single generator is subject to intermittent noise. Generators are controlled by difference control (left), direct control (center), combined control (right) in the extended control topology $c_{ij}^{ext}$. Node index $i$ on the x-axis and time $t$ on the y-axis. Color indicates the frequency $\dot{\vartheta}$. The index $i$ of the affected generator is noted in the upper right corner of each panel. Dashed horizontal lines indicate $t_{start}$ and $t_{end}$, respectively. Parameters as in Fig. \ref{space_time_plot_nocontrol}.
	}
	\label{space_time_plot_singlecontrol_ext}
\end{figure*}

In Fig.\thinspace \ref{fig:CMode_Comp_single}\thinspace (a),\thinspace (b) the ability of the different control schemes to preserve frequency synchronization in the presence of the perturbation is illustrated.
\textit{Difference control} is only effective in counteracting the perturbation if additional links between the generators are present in the communication layer, while \textit{direct control} is only effective in the absence of additional connections.
As detailed in Fig. \ref{space_time_plot_singlecontrol_local}, when applying \textit{difference control} to a sparse control network (i.e. generators in the control layer are connected just to their direct neighbors), the underlying power grid is unable to recover full frequency synchronization after the perturbation ends, since the frequency shift between the northern and the southern parts remains unchanged.
If the generators in the control layer are globally connected (Fig. \ref{space_time_plot_singlecontrol_ext}), full frequency synchronization is always achieved after the end of the perturbation with \textit{difference control}, except if node $i=73$ is perturbed: in this case short-living fluctuations emerge, on a time $t\sim 150$, that possess an intensity of $\Delta \omega \sim 0.2$. 
On the other hand, the application of \textit{direct control} to a local control layer topology (Fig. \ref{space_time_plot_singlecontrol_local}) allows for the achievement of full frequency synchronization at the end of the perturbation. Moreover full frequency synchronization is mostly retained even during the perturbation, irrespectively of short-living fluctuations comprising the whole network that emerge for a time $t\sim 100$. When \textit{direct control} is applied to the extended control layer topology (Fig. \ref{space_time_plot_singlecontrol_ext}), the frequency synchronization is not retained during the perturbation and a frequency shift between the northern and southern parts emerges. Finally, \textit{combined control} is as effective as the more efficient of its two components depending on topology, as the ineffective component is mostly inactive.

\begin{figure}[t]
    \includegraphics[width=1\linewidth]{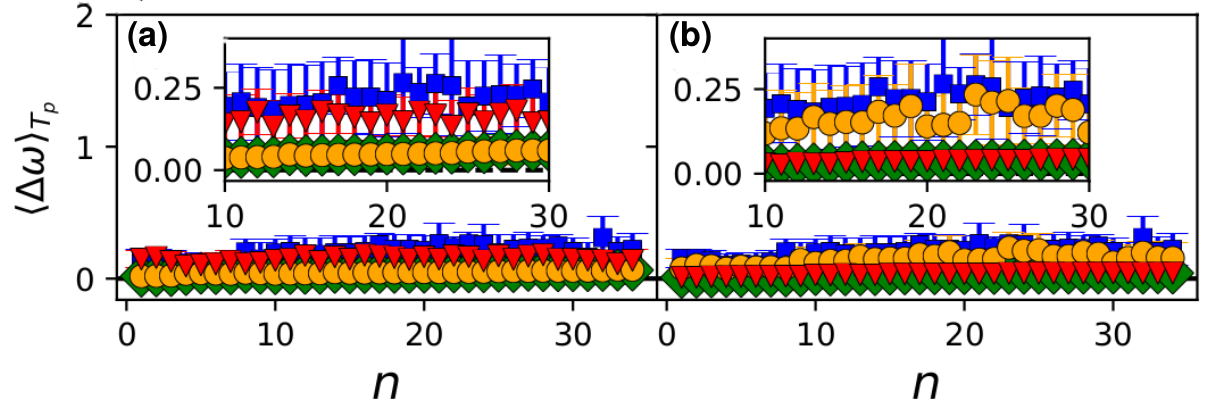}
	\caption{Control of frequency synchronization for intermittent noise targeting multiple nodes: Mean standard deviation of frequency during the time of perturbation $\langle\Delta\omega\rangle_{T_p}$ vs the number $n$ of targeted nodes. The symbols indicate different control schemes. Blue squares: no control. Red triangles: \textit{difference control}. Yellow circles: \textit{direct control}. Green diamonds: \textit{direct control}. Control strength $G=0.04$. The two panels show different control layer topologies: (a) $c_{ij}^{loc}$, (b) $c_{ij}^{ext}$.
    Parameters of the perturbations as in Fig.\thinspace\ref{fig:CMode_Comp_single}. In the inset an enlargement is provided.
	}
	\label{fig:CMode_Comp_multi}
\end{figure}

\begin{figure*}[t]
	\includegraphics[width=0.33\linewidth]{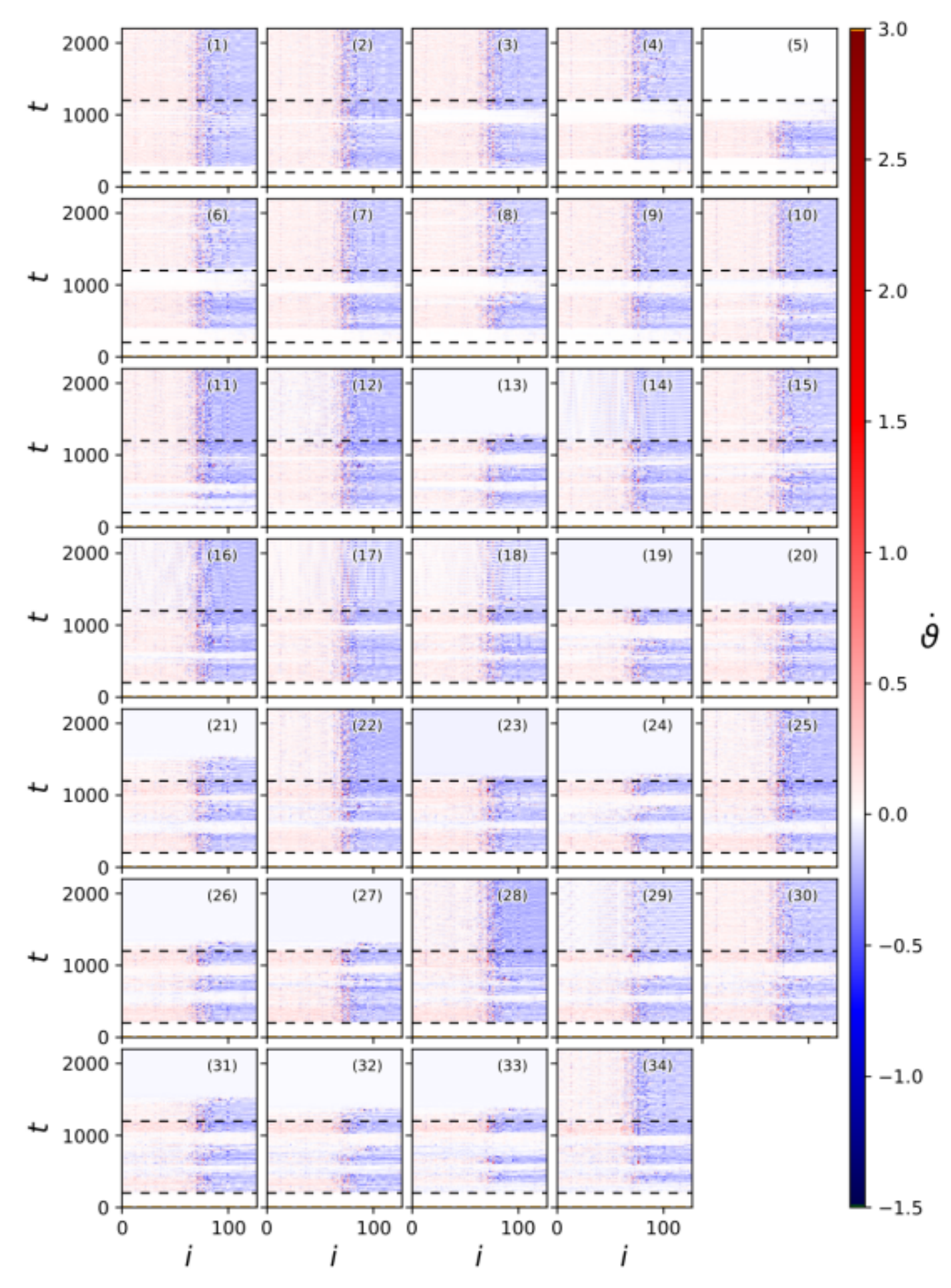}
	\includegraphics[width=0.33\linewidth]{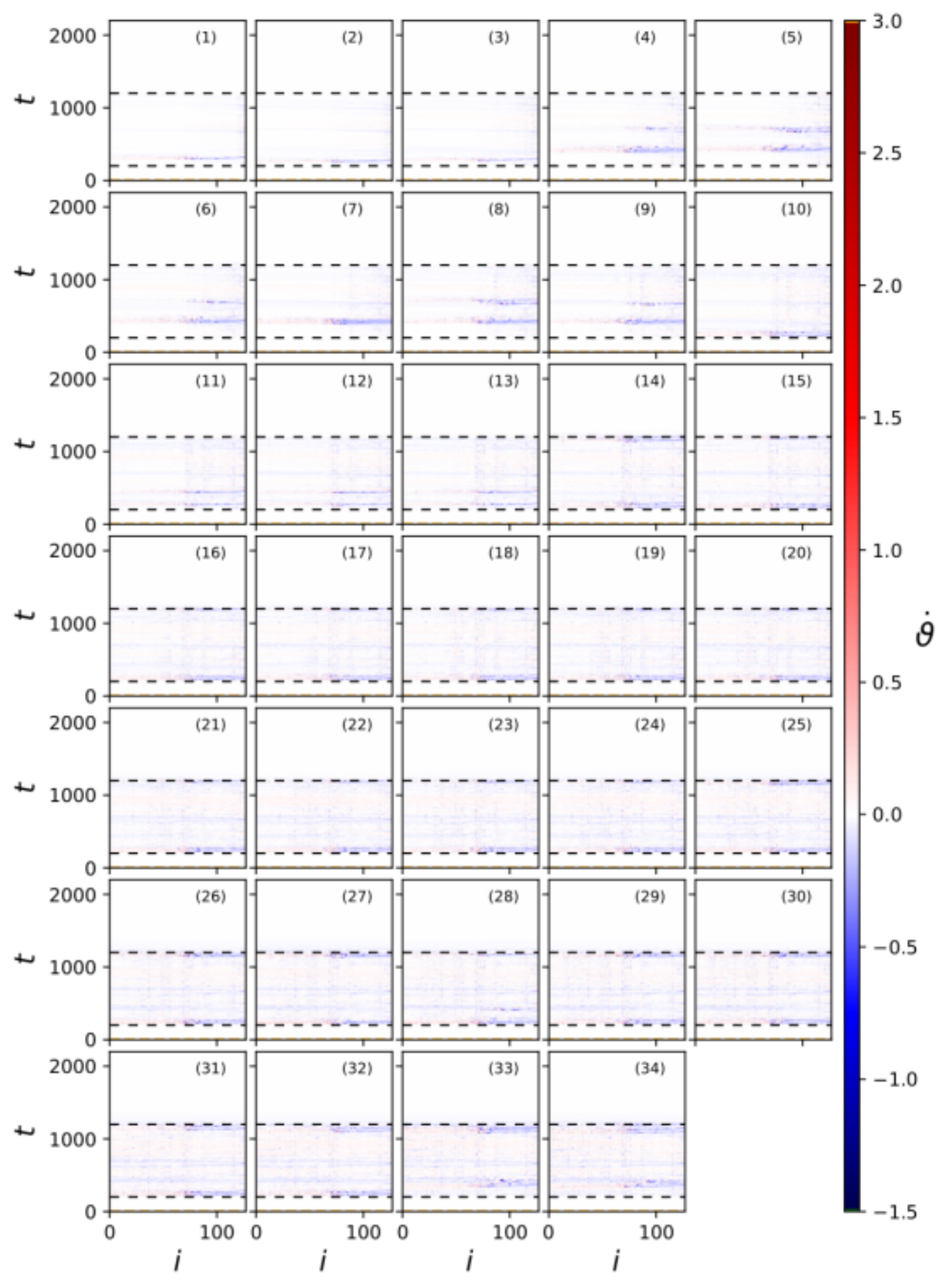}
	\includegraphics[width=0.33\linewidth]{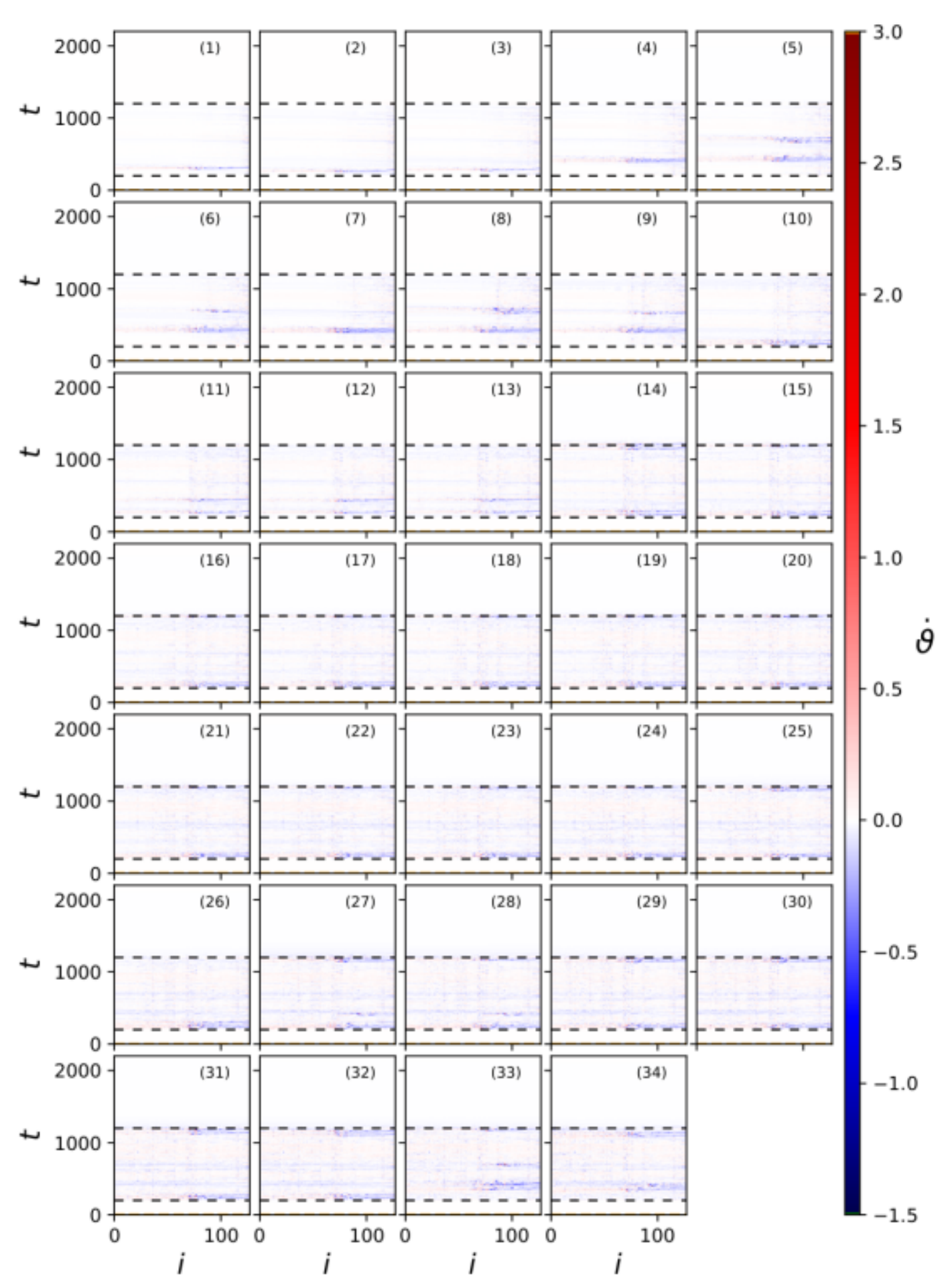}
	\caption{Space-time plots of the network, when multiple generators are subject to intermittent noise. Generators are controlled by difference-control (left), direct-control (center), combination-control (right) in the local control topology $c_{ij}^{loc}$. Node index $i$ on the x-axis and time $t$ on the y-axis. Color indicates the frequency $\dot{\vartheta}$. 
	The number $n$ of the affected generators is noted in the upper right corner of each panel. Dashed horizontal lines indicate $t_{start}$ and $t_{end}$ respectively.
    Parameters as in Fig. \ref{space_time_plot_nocontrol}.
	}
	\label{space_time_plot_multiplecontrol_local}
\end{figure*}

\begin{figure*}[t]
	\includegraphics[width=0.33\linewidth]{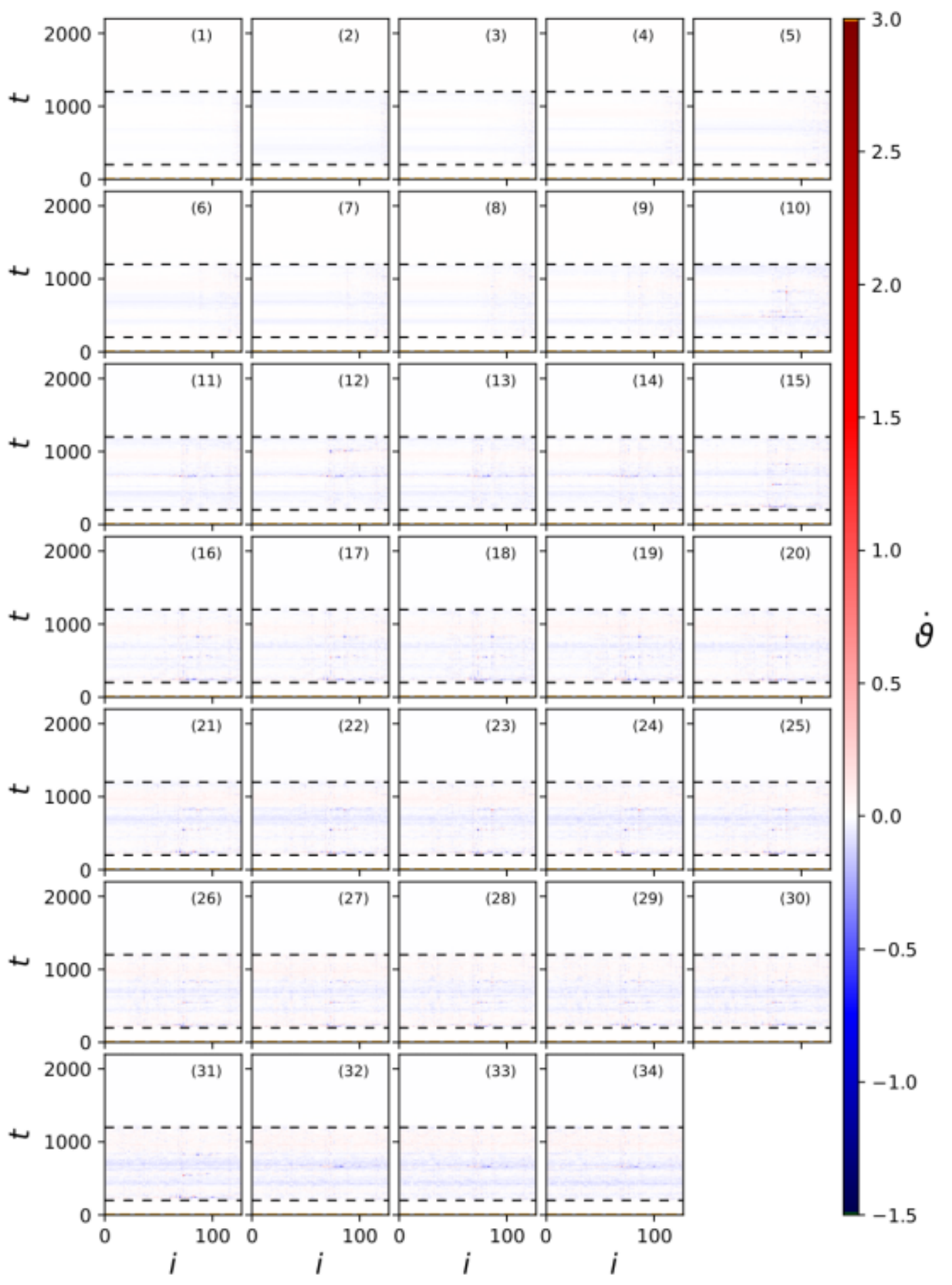}
	\includegraphics[width=0.33\linewidth]{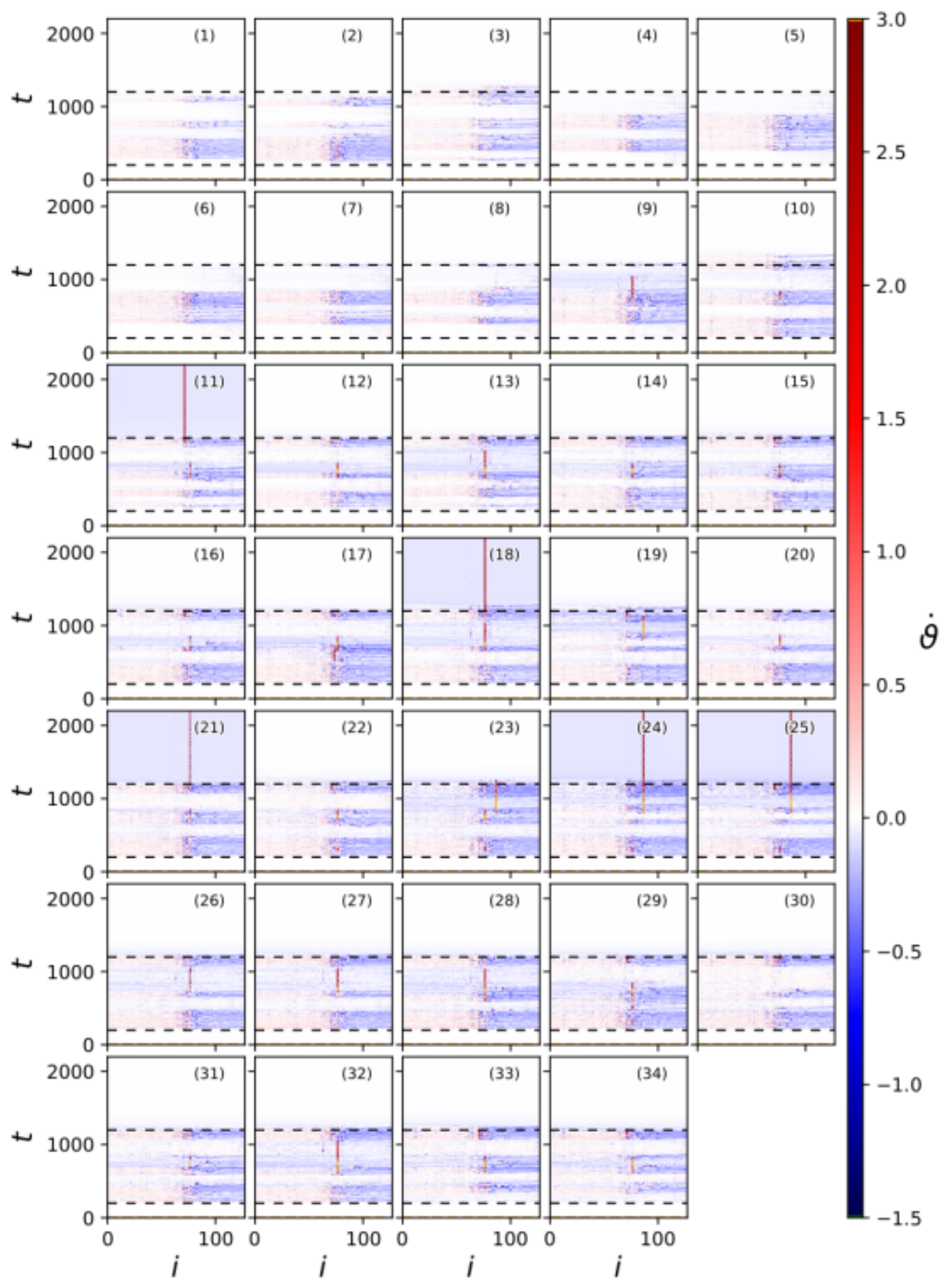}
	\includegraphics[width=0.33\linewidth]{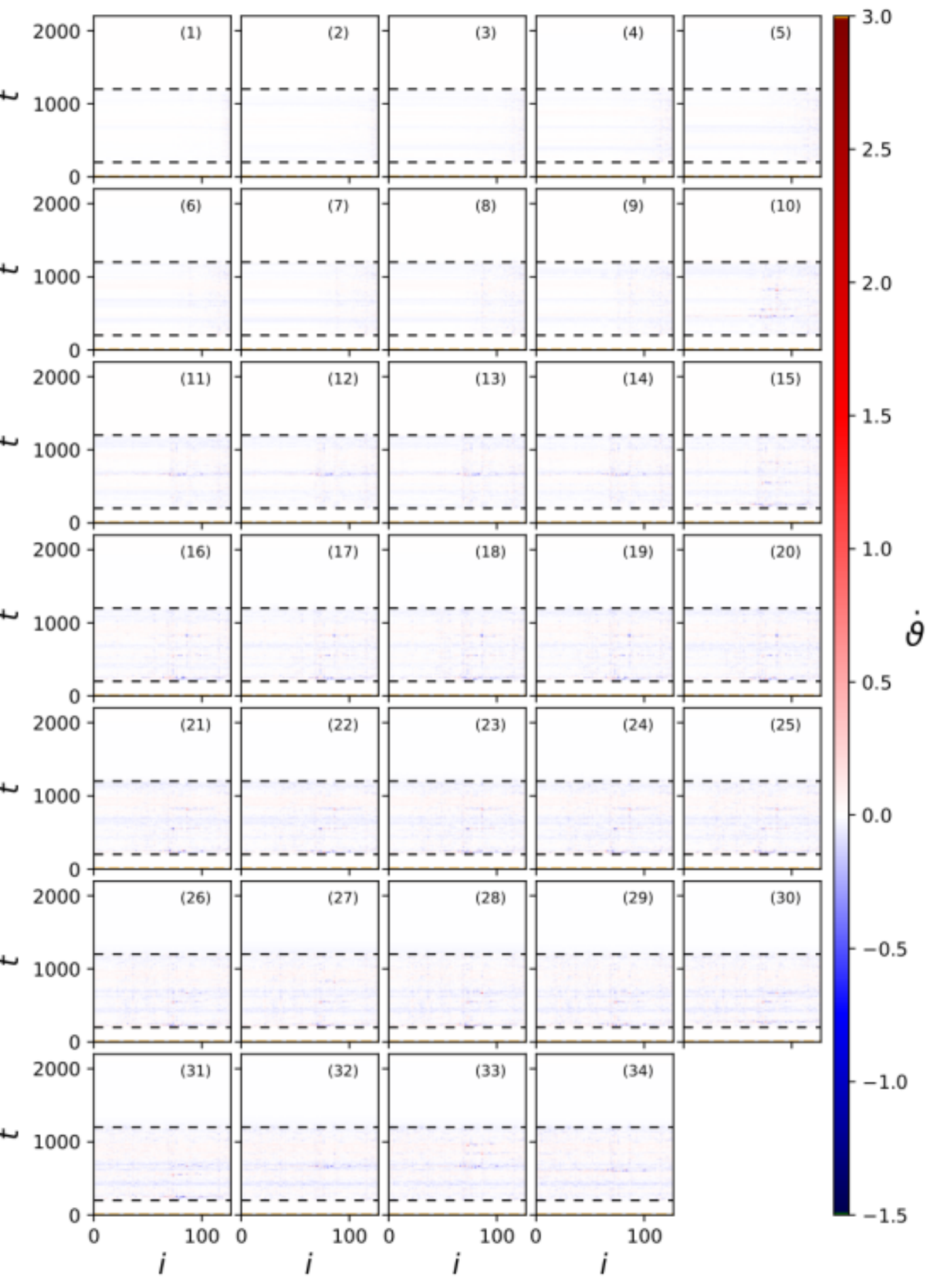}
	\caption{Space-time plots of the network, when multiple generators are subject to intermittent noise. Generators are controlled by difference-control (left), direct-control (center), combination-control (right) in the extended control topology $c_{ij}^{ext}$. Node index $i$ on the x-axis and time $t$ on the y-axis. Color indicates the frequency $\dot{\vartheta}$. The number $n$ of the affected generators is noted in the upper right corner of each panel. Dashed horizontal lines indicate $t_{start}$ and $t_{end}$ respectively. Parameters as in Fig. \ref{space_time_plot_nocontrol}.
	}
	\label{space_time_plot_multiplecontrol_ext}
\end{figure*}

In addition to targeting each individual generator in the network with different perturbations, we now investigate the impact of targeting several generators simultaneously (\textit{multiple node perturbation}). For this purpose we systematically increase the number $n$ of targeted generators. As shown in Fig. \ref{space_time_plot_nocontrol} (right panels), generators are perturbed successively from south to north along the Italian grid: the perturbation first affects nodes in the southern part of the network (characterized by higher node index), and generators with decreasing node index are added successively, one by one, to the list of the perturbed nodes. Thus the perturbation propagates from the south to the north of Italy. 
If intermittent noise is applied to multiple generators, we generally observe a frequency shift between northern and southern parts with stronger fluctuations at the boundary of the two parts (Fig.\thinspace\ref{space_time_plot_nocontrol}). When the perturbation ends, the shift may persist, or single-node desynchronization may occur. In particular if generators close to the boundary are perturbed (i.e., $i=76,71$), for $11\leq n\leq 15$ the fluctuations across the network persist after the end of the perturbation, and a single generator desynchronizes, $i=71$. The remaining network recovers frequency synchronization.
Fig. \ref{fig:CMode_Comp_multi} shows the impact of applying different control schemes to an increasing number of generators subject to intermittent noise.
We observe, in agreement with the case where single generators are targeted, that \textit{direct control} is the most effective control scheme in the absence of additional links between the generators in the control layer (as detailed in Fig. \ref{space_time_plot_multiplecontrol_local}).
However, the frequency deviation induced by the perturbation is the larger, the more nodes are affected simultaneously. Thus the reliability of \textit{direct control} deteriorates with the severity of the perturbation. 
In the case of multiple generators connected in a chain, \textit{direct control} can lead to runaway desynchronization, as the middle generator tries to compensate for its two neighbors which in turn act to compensate for the middle generator. Therefore, if one generator adopts a negative frequency, while the other has a positive frequency, \textit{direct control} will only ensure that their amplitudes are similar, as it only acts to restore the mean frequency in the neighborhood of the controlled node to the nominal frequency of the network.
\textit{Difference control} by itself is ineffective in restoring synchronization during the time of perturbation without additional links between the generators in the communication layer, but it becomes very effective at preserving frequency synchronization within the power grid, if all generators are connected in the communication layer (as detailed in Fig. \ref{space_time_plot_multiplecontrol_ext}).
In the absence of additional links in the communication layer, \textit{combined control} is governed by the interplay of its two components (Figs.~\ref{fig:CMode_Comp_multi}, \ref{space_time_plot_multiplecontrol_local}): the \textit{difference control} is preventing the \textit{direct control} from stabilizing an equilibrium between the generators that is far from frequency synchronization, which greatly improves the effectiveness of merely \textit{direct control}.
In the presence of additional links between the generators in the communication layer (Figs.~\ref{fig:CMode_Comp_multi}, \ref{space_time_plot_multiplecontrol_ext}) the dynamics of \textit{direct control} is again mostly dominated by its \textit{difference control} component.

\section{Conclusion}
The control of intermittent energy fluctuations in power grids is of great current interest in view of the impending termination of nuclear and fossil-fuel energy and replacement of old power plants with renewable energy sources \cite{UNF15}. Wind and photovoltaic power are the most promising technologies, but their integration into the grid poses a challenge \cite{JAC11,TUR99,UEC15}, in particular due to the fluctuation features of renewable power generators \cite{Anvari2016,ANV17a,AUE17,Schmietendorf:EPJB:90} and the impact of energy trading on the network \cite{SCH18c,RIT21}.
In this work, we have investigated the controllability of a power grid against intermittent noise using the Italian grid as proof-of-principle. 
We have presented a novel approach by considering the dynamics of a power grid in a two-layer network model, using a fully dynamical description for the communication layer. 
Specifically, here we have modelled the Italian high voltage power grid as a dynamical two-layer network, where the dynamics of the power grid layer is described in terms of the second order Kuramoto model with inertia. On the other hand the second layer, which represents the communication network, models the dynamic control signal for each generator.
To describe the fluctuating power output of renewable energy power plants realistic intermittent noise has been used.
Previous investigations of multiple-layer power grids have been performed by taking into account only static nodes without dynamics, focusing on topological effects \cite{BUL10}. On the other hand, investigations of the dynamics of the (Italian) power grid are usually conducted only in a single layer \cite{serrani2004,fortuna2012,olmi2014,TUM19,mehrmann2018,TAH19}, as well as the investigation of the dynamics of cascading failures \cite{SCH18z}.
\\
In the communication layer we have investigated a selection of different control schemes (control functions $f_i^{diff}$, $f_i^{dir}$ and $f_i^{comb}$) and control topologies (adjacency matrices $c_{ij}^{loc}$ and $c_{ij}^{ext}$). All control schemes take advantage of the second layer by collecting information from adjacent nodes described by $c_{ij}$ to calculate the control signal. This can be done either in a local setting ($c_{ij}^{loc}$) where generators possess the same communication links as in the power grid layer, or in an extended control layer topology ($c_{ij}^{ext}$) where additional communication links between all generators are present. We have tested (i) a control scheme aimed at synchronizing the frequency of the controlled nodes with their neighbors (difference control $f^{diff}$), (ii) a control scheme aimed at restoring the original synchronization frequency in the neighborhood of the controlled node (direct control $f^{dir}$), and (iii) a mixed approach combining both ($f^{comb}$). 
\textit{Difference control} proves ineffective in the local control topology $c_{ij}^{loc}$, because it is only able to improve frequency synchronization locally. 
This means that if a perturbation causes the northern and southern part of the network to lose synchronization, the control scheme only suppresses local fluctuations in the northern and southern part separately, but does not restore frequency synchronization between the two parts of the power grid: no generator is sufficiently connected to both parts simultaneously to make the control scheme effective. 
However, this shortcoming is removed when we consider the extended control topology.
In particular, when additional communication links are introduced in $c_{ij}^{ext}$, all generators become well connected to both the northern and southern part of the grid, enabling the control to restore frequency synchronization across the whole grid. 
\textit{Direct control} proves to be more effective in the local control topology $c_{ij}^{loc}$, where less communication links are present than in $c_{ij}^{ext}$. 
This is due to the basis mechanism underlying \textit{direct control}: it compensates the deviation of the mean frequency of all nodes connected to the controlled generator, thus causing the control to remain inactive when the mean frequency matches the nominal frequency of the power grid, while not all nodes are frequency synchronized.
Adding further communication links in the control topology renders the control scheme ineffective as multiple controlled generators compensate each other instead of restoring the nominal frequency 
within the power grid.
\textit{Combined control} in the local control topology $c_{ij}^{loc}$ is governed by its \textit{direct control} component. Since synchronization is lost across the grid, but not locally, \textit{difference control} is mostly inactive, while the \textit{direct control} part is responsible for restoring synchronization within the grid. 
When additional communication links between the generators are present (i.e. $c_{ij}^{ext}$), the combined control is dominated by its \textit{difference control} component. In this case the \textit{direct control} is mostly inactive since the mean frequency across the two desynchronized regions of the network is equal to the nominal frequency. 
\\
The investigation of the self-emerging control dynamics following perturbations has highlighted the role played by some specific nodes: dead-ends and dead-trees result to be always problematic, in agreement with recent work \cite{MEN14,AUE17} where it has been demonstrated that the cost-minimizing creation of dead-end or dead-tree structures increases the vulnerability of the power grid to large perturbations. The role of solitary nodes has recently been emphasized~\cite{TAH19,HEL20,BER21a}. Moreover, it turns out that the Italian power grid can be divided in two specific parts: the northern, continental part, with a higher average connectivity, which is more resilient to perturbations, and the southern, peninsular part, characterized by a low average connectivity. The elongated structure of the southern part makes it less robust to perturbations. These results are in agreement with previous findings by \cite{TOT20}, where also other realistic perturbation scenarios were applied.
\\
Further work should incorporate time delay in the dynamics of the control grid, accounting for the delay in detecting the frequency of the generators and the delays during communication between them.
Furthermore it would be interesting to investigate different topologies in the communication layer, which may be distinct from the underlying power grid topology (such as random networks).

\section*{Acknowledgements}
The authors acknowledge Enrico Steinfeld and Katrin Schmietendorf for useful discussions. 
Funded by the Deutsche Forschungsgemeinschaft (DFG, German Research Foundation) - Projektnummer 163436311 - SFB 910. 

\bibliography{biblio.bib}


\end{document}